\documentclass[times,twocolumn,final]{elsarticle}

\usepackage{cnf}
\makeatletter\gdef\cnf@pagecountlimit{9999}\makeatother
\biboptions{sort&compress,numbers}
\usepackage{framed,multirow}

\usepackage{amssymb}
\usepackage{latexsym}
\usepackage{amsmath}
\usepackage{subcaption}
\usepackage{siunitx}
\DeclareSIUnit\bar{bar}
\usepackage[version=4]{mhchem}

\usepackage{url}
\usepackage{xcolor}
\definecolor{newcolor}{rgb}{.8,.349,.1}

\usepackage{hyperref}

\begin{document}

\begin{frontmatter}

\title{Efficient sensitivity analysis for combustion problems via reformulation and adjoint methods}%

\author[]{Francesca Loffredo\corref{cor1}}
\tnotetext[cor1]{Corresponding author}
\emailauthor{f.loffredo@itv.rwth-aachen.de}{F. Loffredo}
\author[]{Raymond Langer}
    
\author[]{Marco Vivenzo}
\author[]{Sanket Girhe}
\author[]{Heinz Pitsch}

\address{Institute for Combustion Technology, RWTH Aachen University, Templergraben 64, 52056 Aachen, Germany}

\begin{abstract}
Numerical simulations of combustion processes rely on chemical kinetic models, whose predictive capability is typically assessed by comparing measurements of quantities of interest~(QoIs) in idealized experimental configurations with the corresponding simulated values. Sensitivity analysis quantifies the influence of model parameters on QoIs, making it a key tool for model development and optimization.
Given the large number of parameters involved in detailed kinetic models compared to the number of QoIs, adjoint methods are a promising approach for efficient sensitivity analysis. 
Ignition delay times~(IDTs) and extinction strain rates~(ESRs) are among the QoIs of primary importance. However, applying adjoint methods to these quantities is not straightforward and is addressed in this work.
The system of governing equations is reformulated via a coordinate transformation that makes the QoI an explicit component of the solution vector. This reformulation enables the application of adjoint sensitivity analysis for both IDT and ESR.
A computational complexity analysis demonstrates that the proposed formulation significantly improves algorithmic efficiency compared to conventional approaches.
The methodology is verified by comparing adjoint sensitivity coefficients with those obtained from other methods.
Execution time measurements show computational cost reductions of up to two orders of magnitude relative to standard direct sensitivity analysis.
A benchmark study is conducted across kinetic models of increasing size to assess how far theoretical efficiency gains are realized in practice. While the results show that implementation details can limit performance for smaller problem sizes, the expected asymptotic scaling is observed for larger models. The benchmark thus provides a reference framework for identifying key performance bottlenecks and guiding subsequent optimization efforts.

\end{abstract}

\begin{keyword}
\KWD Kinetic model \sep Sensitivity analysis\sep Adjoint \sep Ignition delay time\sep Extinction strain rate \sep Continuation methods 
\end{keyword}

\end{frontmatter}

\section{Introduction}
\label{section:Introduction}
The development and validation of chemical kinetic models rely on the comparison of numerically predicted quantities of interest~(QoIs) with corresponding measurements from well-controlled experimental configurations. Among these, ignition delay time~(IDT) and extinction strain rate~(ESR) represent some of the fundamental QoIs. The IDT is a marker of the autoignition propensity of a fuel and is therefore a critical design parameter in applications where precise control of ignition timing affects performance, efficiency, and pollutant formation~\cite{Miron2021, Mustayen2022}. The ESR provides fundamental insights into the stability limits of premixed and non-premixed flames. Given their relevance in practical applications, the development of kinetic models that accurately predict IDTs and ESRs is central to the advancement of combustion technologies.

To achieve high predictive accuracy, targeted refinement of the most influential model parameters is often performed to improve a model's agreement with experimental observations. These critical parameters can be identified through sensitivity analysis, which quantifies the influence of kinetic, thermodynamic, and transport parameters on QoI predictions~\cite{turanyi2014analysis, Tomlin2013}.

Several approaches~\cite{turanyi2014analysis} have been proposed for computing sensitivity coefficients. Here, we focus on methods for evaluating local, first-order sensitivity coefficients, which have been widely used to quantify the influence of model parameters on combustion-relevant QoIs. 
The brute-force method consists of performing an initial baseline simulation followed by a series of additional simulations in which individual input parameters are perturbed one at a time. The sensitivity coefficients for a given parameter are then computed from the change in QoI between the baseline and the corresponding perturbed simulation. 
Although the brute-force method is particularly appealing for its flexibility in computing sensitivity coefficients for any QoI, it becomes computationally prohibitive for large models.
In addition, a suitable and potentially parameter-specific~\cite{Long2019b} magnitude of the perturbation factor must be selected, as it must balance round-off errors at low perturbation and nonlinear effects at high perturbation that affect the validity of the local linear sensitivity approximation.
More recently, another route has been proposed, in which sensitivity coefficients are obtained from an eigenvalue analysis of the linearized kinetic system~\cite{Hu2025}. Although this approach may offer a more computationally efficient alternative to the brute-force method, the accuracy of the computed sensitivity coefficients depends on the validity of the problem-specific assumptions on which it is based.

Alternatively, sensitivity coefficients can be computed as a direct solution to a system of differential equations~\cite{Ji2019, Langer2021, Reuven1986}.
These approaches, commonly referred to as direct methods, generally require a mathematical problem formulation in which the quantities of interest are encoded as scalar-valued objective functions that explicitly depend on system outputs. This requirement has led some authors~\cite{Long2019b, Ji2019, Lemke2019, Almohammadi2021} to use approximations for IDTs and ESRs in direct sensitivity analyses, since these quantities are generally not expressed as explicit functions of the system outputs. 
Ji~et al.~\cite{Ji2019} demonstrated a strong correlation between the temperature sensitivity at the ignition point and the brute‑force sensitivity of the IDT, and therefore proposed to approximate the computationally expensive brute‑force IDT sensitivity analysis with a less demanding direct method applied to the temperature at the ignition point.
Similarly, Ji~et al.~\cite{Ji20202} suggested analyzing laminar flame speed sensitivities obtained via direct methods as a computationally efficient alternative to brute-force computation of premixed flames' ESR sensitivities.
Nevertheless, potential deviations between the sensitivity coefficients computed for approximate QoI representations and those obtained for the exact QoI cannot be quantified \textit{a priori}, thus always necessitating further validation to assess the viability of the approximation employed~\cite{Long2019b, Lemke2019, Almohammadi2021}. Furthermore, qualitative agreement has been observed between the sensitivity coefficients obtained for approximate and true quantities, mainly for the kinetic rate parameters~\cite{Long2019b, Ji20202}; transport and thermochemistry parameters were not considered, and the suggested approximation may be less accurate. 

To overcome these challenges and enable the application of conventional direct sensitivity analysis methods (hereafter referred to as forward methods), several authors~\cite{Dong2005, Wang2009, Holley2009, dsa_2019} have reformulated the system of governing equations to incorporate IDT and ESR directly as components of the respective solution vectors.
Once conventional direct sensitivity analysis methods become applicable, combining them with adjoint formulations offers an attractive strategy for improving computational efficiency~\cite{gardner2022sundials, hindmarsh2005sundials}. In particular, the adjoint approach becomes increasingly advantageous when the number of parameters is large, owing to its superior algorithmic efficiency~\cite{Braman2015, Langer2021}. However, the combination of these reformulation strategies with adjoint-based methods remains unexplored. Therefore, it constitutes the primary focus of this study.

In this work, IDT and ESR are incorporated as explicit components of the solution vector by recasting the system of governing equations through suitable coordinate transformations~\cite{dsa_2019, Giovangigli1989}. This formulation enables the derivation of adjoint sensitivity coefficients for the reformulated system. Building on this, the method is applied to derive sensitivity coefficients for problems involving up to 30\,000 parameters.
The resulting sensitivity coefficients are compared with those obtained using the forward and brute-force methods, and the implementation is further benchmarked for kinetic models of increasing sizes. Execution-time measurements are collected to quantify the performance improvement of the adjoint method and compared with its theoretical predictions to assess the extent to which the algorithmic efficiency gains of the adjoint approach can be realized in practice.

\section{Methodology}\label{section:Methodology}
This section introduces the mathematical and numerical framework used to derive adjoint sensitivity coefficients for both IDT and ESR. These QoIs, typically measured in well-controlled experimental configurations, are numerically obtained by solving a set of governing equations that model idealized systems. 
For both QoIs, the respective governing equations, the coordinate transformation incorporating the QoI into the system solution vector, and the derivation of adjoint sensitivity coefficients for the reformulated systems are presented.
\subsection{Ignition delay time}
Experimental configurations used to measure IDTs are typically modeled as closed, adiabatic, homogeneous systems.  
Under these assumptions, and considering an ideal gas mixture, the governing equations reduce to a set of unsteady, zero-dimensional ordinary differential equations~(ODEs) formulated as an initial value problem~(IVP)
\begin{align}
\begin{split}
\label{eq:ivp} 
&\frac{\mathrm{d} Y_k}{\mathrm{d} t} = \frac{\dot{\omega}_k M_k}{\rho},  \quad k = 1,\,\dots,\,n_{\mathrm s} \,, \\[6pt]
&\frac{\mathrm{d} T}{\mathrm{d} t} = -\frac{1}{\rho c_v} \left( \sum_{k=1}^{n_{\mathrm s}} \left( h_k M_k - R T \right)\dot{\omega}_k - \frac{p}{\rho} \frac{\mathrm{d}\rho}{\mathrm{d} t} \right),
\end{split}
\end{align}
where \( t \) denotes time and \( n_{\mathrm s} \) is the number of species in the kinetic model.  Here, \( \rho \), \( T \), and \( p \) denote the mixture density, temperature, and pressure, respectively; \( R \) is the universal gas constant, and \( c_v \) is the mixture-averaged specific heat capacity at constant volume. The quantities \( Y_k \), \( M_k \), \( \dot{\omega}_k \), and \( h_k \) represent the mass fraction, molar mass, production rate, and specific enthalpy of species \( k \), respectively. Under isochoric or isobaric conditions, the temperature equation in Eq.~\eqref{eq:ivp} can be simplified.
The system of equations is closed by the ideal gas law and the initial conditions at \( t_0 = 0 \),
\begin{align}
\begin{split}
\label{eq:y_hr}
Y_k(t_0) = Y_{k,0}, \quad k = 1,\,\dots,\,n_{\mathrm s},  \quad T(t_0) = T_0. 
\end{split}
\end{align}

Together with the vector of input model parameters \( \boldsymbol{\alpha} \), the IVP defined by Eqs.~\eqref{eq:ivp} and~\eqref{eq:y_hr} can be expressed as
\begin{equation}
\label{eq:ivp_compact}
    \frac{\mathrm{d} \boldsymbol{y}^{\mathrm{hr}}}{\mathrm{d} t} = \boldsymbol{F}^{\mathrm{hr}}\big(t,\boldsymbol{y}^{\mathrm{hr}}(t), \boldsymbol{\alpha}\big), \quad \boldsymbol{y}^{\mathrm{hr}}(t_0) = \boldsymbol{y}^{\mathrm{hr}}_0,
\end{equation}
where \( \boldsymbol{y}^{\mathrm{hr}}(t) \) is the homogeneous-reactor solution vector defined as
\begin{equation}
\boldsymbol{y}^{\mathrm{hr}}(t) = \big(Y_1(t),Y_2(t),\,\dots,\,Y_{n_{\mathrm s}}(t),\, T(t)\big) \,.
\end{equation}

The IDT is typically defined as the time at which a function of the dependent variables or their derivatives satisfies a prescribed criterion. It is therefore an implicitly defined quantity of the system solution.
As a result, the perturbative information required to compute sensitivity coefficients cannot be obtained directly from the system Jacobian, in contrast to explicit functions of the system solution.
To overcome this limitation, the coordinate transformation proposed by Gururajan and Egolfopoulos~\cite{dsa_2019} is employed. The dimensionless time variable \( t' = t / \tau \) is introduced so that the initial value problem in Eq.~\eqref{eq:ivp_compact} is reformulated into a two-point boundary value problem, where \(\tau\) represents the ignition delay time. The reformulated system reads
\begin{align}
\begin{split}
\label{eq:ivp_recast}
 & \dfrac{\mathrm{d} \boldsymbol{y}^{\mathrm{hr}}}{\mathrm{d} t'} = \tau\, \boldsymbol{F}^{\mathrm{hr}}\big(\tau t',\boldsymbol{y}^{\mathrm{hr}}(\tau t'), \boldsymbol{\alpha} \big), \\[6pt]
 &\dfrac{\mathrm{d} \tau}{\mathrm{d} t'} = 0,
\end{split}
\end{align}
with boundary conditions given as
\begin{align}
\begin{split}
t' = 0: &\quad \boldsymbol{y}^{\mathrm{hr}} - \boldsymbol{y}_0^{\mathrm{hr}}= 0, \\
t' = 1: &\quad T - T_{\mathrm{ign}} = 0,
\end{split}
\end{align}
where \(T_{\mathrm{ign}}\) is the temperature at the ignition point. 
The trivial equation for the IDT in Eq.~\eqref{eq:ivp_recast} is introduced so that the ignition delay time is treated as an eigenvalue to be determined, requiring the specification of the additional boundary condition on temperature at the right boundary. Consequently, the ignition delay time appears explicitly as a dependent variable.

Discretizing Eqs.~\eqref{eq:ivp_recast} with a first-order backward-difference scheme for the species and temperature equations and a forward-difference scheme for the equation for \(\tau\) yields the nonlinear algebraic system
\begin{equation}
\label{eq:hrbvp_recast}
\hat{\boldsymbol{F}}^{\mathrm{hr}}\left(\hat{\boldsymbol{y}}^{\mathrm{hr}},\boldsymbol{y}^{\mathrm{hr}}_{0},T_{\mathrm{ign}},\boldsymbol{\alpha}\right)=0,
\end{equation}
where \(\hat{\boldsymbol{F}}^{\mathrm{hr}}\) denotes the residual of the discretized reformulated system.
The solution vector of the augmented system is defined as
\begin{equation}
\label{eq:y_hr_hat}
\hat{\boldsymbol{y}}^{\mathrm{hr}}(t^{\prime}) 
= \left(\boldsymbol{y}^{\mathrm{hr}}(t^\prime), \tau \right)^{\top}.
\end{equation}
The IVP in Eq.~\eqref{eq:ivp_compact} is first solved. Then, the resulting normalized-time grid and solution values are adopted as the initial guess for the augmented problem, which is solved using Newton's method.
\subsection{Extinction strain rate}

Extinction strain rates are determined experimentally by progressively increasing the opposing flow rates in counterflow flames until extinction occurs, while maintaining balanced axial momentum.
The momentum balance is typically enforced by imposing the same momentum flux \(M\) at both sides
\begin{equation} 
\label{eq:momentum}
M = \rho_{\mathrm l} v_{\mathrm l}^{2} = \rho_{\mathrm r} v_{\mathrm r}^{2}\,, 
\end{equation}
where \(\rho_{\mathrm l}\) and \(v_{\mathrm l}\) denote the density and axial velocity on the left side, and \(\rho_{\mathrm r}\) and \(v_{\mathrm r}\) are the corresponding properties on the right side. 

Taking into account the momentum balance and following the analysis by Seshadri and Williams~\cite{seshadri1978}, the global strain rate is defined as
\begin{equation}
\label{eq:strain_rate_sheshadri}
a = \frac{4}{L} \sqrt{M / \rho_{\mathrm r}},
\end{equation}
where \(L\) denotes the separation distance between the burner inlets. This definition is adopted throughout the remainder of this work to characterize counterflow flames and extinction conditions.
In numerical simulations, natural continuation methods~\cite{keller1977}  follow an analogous procedure to predict ESR. Starting from a stable flame solution, progressively increasing values of \(M\) are imposed, thereby increasing the mass fluxes at the boundaries. The governing equations are solved with the incrementally changing boundary conditions until the extinction point is reached. These governing equations are given by~\cite{kee89_counterflow}
\begin{align}
\begin{split}
\label{eq:bvp}
&\frac{\mathrm{d}(\rho v)}{\mathrm{d} y} + 2\rho G = 0 \, , \\[6pt]
&\rho G^2 + \rho v \frac{\mathrm{d} G}{\mathrm{d} y} - \frac{\mathrm{d}}{\mathrm{d} y}\left(\mu \frac{\mathrm{d} G}{\mathrm{d} y}\right) + p' = 0 \, , \\[6pt]
&\rho v \frac{\mathrm{d} Y_k}{\mathrm{d} y} + \frac{\mathrm{d} j_k}{\mathrm{d} y} - \dot{\omega}_k M_k = 0, \quad k = 1,\,\dots,\,n_{\mathrm s} \, , \\[6pt] 
&\rho v c_p \frac{\mathrm{d} T}{\mathrm{d} y} - \frac{\mathrm{d}}{\mathrm{d} y}\left(\lambda \frac{\mathrm{d} T}{\mathrm{d} y}\right) + \dot{q}_{\mathrm{R}} 
 + \\ & \quad\quad + \sum_{k=1}^{n_{\mathrm s}} j_k c_{p,k} \frac{\mathrm{d} T}{\mathrm{d} y}
 + \sum_{k=1}^{n_{\mathrm s}} h_k \dot{\omega}_k M_k = 0 \, , \\[6pt]
\end{split}
\end{align}
where \(y\) denotes the spatial coordinate along the axial direction, \(v\) the axial velocity, and \(G = G(y)\) the radial gradient of the radial velocity. The parameters \(\lambda\), \(\mu\), and \(c_p\) represent the mixture thermal conductivity, dynamic viscosity, and specific heat at constant pressure, respectively. The species flux and specific heat at constant pressure of species \(k\) are denoted by \(j_k\) and \(c_{p,k}\), respectively, while the net radiative heat flux, \(\dot{q}_{\mathrm{R}}\), is computed under the optically thin transfer assumption~\cite{Barlow2001}. Transport processes are modeled using a mixture-averaged transport model that includes the Soret effect~\cite{LangerPhd}. The reduced radial pressure gradient, \(p' = -(1/x)\,\mathrm{d}p/\mathrm{d}x\), where \(x\) denotes the radial coordinate, is introduced as an eigenvalue to ensure that the system is well-posed when the inlet mass fluxes are prescribed on both sides.
The system is closed by imposing boundary conditions at the left
(\(y = 0\)) and right (\(y = L\)) sides of the domain, given by
\begin{align}
\begin{split}
\label{eq:cfbvp_boundaries}
y = 0:\quad
& \rho v - \sqrt{\rho_{\mathrm l} M} = 0\,, \\
& G = 0, \\
& \rho v (Y_k - Y_{k,\mathrm l}) + j_k = 0 \, , \quad k = 1,\,\dots,\,n_{\mathrm s} \, , \\
& T - T_{\mathrm l} = 0\,, \\
y = L:\quad
& \rho v + \sqrt{\rho_{\mathrm r} M} = 0, \\
& G = 0, \\
& \rho v (Y_k - Y_{k,\mathrm r}) + j_k = 0 \, , \quad k = 1,\,\dots,\,n_{\mathrm s} \, , \\
& T - T_{\mathrm r} = 0\,,
\end{split}
\end{align}
where \(Y_{k,\mathrm l}\) and \(T_{\mathrm l}\) denote the mass fraction of
species \(k\) and the temperature imposed at the left boundary,
respectively, and the corresponding quantities with subscript \(\mathrm r\)
refer to the right boundary. In the counterflow diffusion flame
configuration considered here, the left boundary typically corresponds to
the fuel stream and the right boundary to the oxidizer stream, although the same formulation also accommodates premixed streams, in which each boundary may carry a fuel--oxidizer mixture.
For the counterflow flame configuration, the solution vector is
\begin{equation}
\label{eq:y_cf}
\boldsymbol{y}^{\mathrm{cf}}(y) = \big(\rho v(y),  G(y), Y_1(y), \ldots, Y_{n_{\mathrm s}}(y),T(y), p'\big)^{\top}.
\end{equation}
It contains the spatially evolving dependent variables together with the eigenvalue \(p'\), which is constant throughout the domain. 

The family of solutions obtained by solving the system in Eqs.~\eqref{eq:bvp} and~\eqref{eq:cfbvp_boundaries} for varying values of \( M \) defines a curve which, in implicit form, can be written as
\begin{equation}
\label{eq:cfbvp_compact}
\boldsymbol{F}^{\mathrm{cf}}\!\left(\boldsymbol{y}^{\mathrm{cf}}(y),M,\boldsymbol{y}^{\mathrm{cf}}_{\mathrm{l}},\boldsymbol{y}^{\mathrm{cf}}_{\mathrm{r}},\boldsymbol{\alpha}\right)=0\,.
\end{equation}
Here, \( \boldsymbol{y}^{\mathrm{cf}}_{\mathrm{l}} \) and \( \boldsymbol{y}^{\mathrm{cf}}_{\mathrm{r}} \) denote the prescribed species mass fractions and temperature values at the left and right
boundaries, respectively. These values, together with the vector of input parameters \( \boldsymbol{\alpha} \), are held constant while solving Eq.~\eqref{eq:cfbvp_compact}.

Three aspects complicate the computation of sensitivity coefficients for the ESR. First, the QoI is defined through the input variable \( M \) in Eq.~\eqref{eq:strain_rate_sheshadri}, rather than as an explicit function of the solution. Second, the extinction condition corresponds to a turning point on the solution curve,  at which the Jacobian becomes singular, leading to numerical difficulties when solving Eq.~\eqref{eq:bvp} with Newton's method.
Third, because the extinction point is not known \textit{a priori}, it might not be reached by prescribing progressively larger values of \(M\), as any prescribed value may already exceed the extinction limit.

These difficulties are addressed by parameterizing the solution curve by the arc length \( s \), yielding
\begin{equation}
\label{eq:F_param}
\boldsymbol{F}^{\mathrm{cf}}\!\left(\boldsymbol{y}^{\mathrm{cf}}(y,s),M(s),\boldsymbol{y}^{\mathrm{cf}}_{\mathrm{l}},\boldsymbol{y}^{\mathrm{cf}}_{\mathrm{r}},\boldsymbol{\alpha}\right)=0\,.
\end{equation}
Differentiating with respect to \( s \) yields the relation
\begin{equation}
\label{eq:cf_tangent}
\frac{\partial\boldsymbol{F}^{\mathrm{cf}}}{\partial\boldsymbol{y}^{\mathrm{cf}}}
\frac{\mathrm{d}\boldsymbol{y}^{\mathrm{cf}}}{\mathrm{d}s}
+
\frac{\partial\boldsymbol{F}^{\mathrm{cf}}}{\partial M}
\frac{\mathrm{d}M}{\mathrm{d}s}
=
\boldsymbol{0}\,,
\end{equation}
where the augmented solution vector is defined by
\begin{equation}
\label{eq:y_cf_hat}
\hat{\boldsymbol{y}}^{\mathrm{cf}}(y,s)
=
\left(
\boldsymbol{y}^{\mathrm{cf}}(y,s)^{\top},\,
M(s)
\right)^{\top}\,,
\end{equation}
with the associated tangent vector to the solution curve
\begin{equation}
\label{eq:cf_tangent_aug}
\frac{\mathrm{d}\hat{\boldsymbol{y}}^{\mathrm{cf}}}{\mathrm{d}s}
=
\left(
\frac{\mathrm{d}\boldsymbol{y}^{\mathrm{cf}}}{\mathrm{d}s},\,
\frac{\mathrm{d}M}{\mathrm{d}s}
\right)^{\!\top}\,.
\end{equation}
Geometrically, the extinction point corresponds to a turning point of the solution branch, where \(M\) attains its maximum and the branch folds back. At this point, the component along \(M\) of the solution curve tangent vanishes, \(\mathrm{d}M/\mathrm{d}s = 0\), while the components along \(\boldsymbol{y}^{\mathrm{cf}}\) are nonzero.
Substituting \(\mathrm{d}M/\mathrm{d}s = 0\) into Eq.~\eqref{eq:cf_tangent} yields
\begin{equation}
\frac{\partial\boldsymbol{F}^{\mathrm{cf}}}{\partial\boldsymbol{y}^{\mathrm{cf}}}
\frac{\mathrm{d}\boldsymbol{y}^{\mathrm{cf}}}{\mathrm{d}s}
=
\boldsymbol{0}\,,
\end{equation}
at the extinction point, with \(\mathrm{d}\boldsymbol{y}^{\mathrm{cf}}/\mathrm{d}s \neq \boldsymbol{0}\). A nonzero vector can satisfy this relation only if the Jacobian \(\partial\boldsymbol{F}^{\mathrm{cf}}/\partial\boldsymbol{y}^{\mathrm{cf}}\) is singular, and hence rank-deficient by at least one. Assuming the turning point is simple~\cite{Calvetti2000,keller1987}, i.e., that it has a single null direction, the rank deficiency is exactly one and the null space is spanned entirely by the tangent direction \(\mathrm{d}\boldsymbol{y}^{\mathrm{cf}}/\mathrm{d}s\). Since \(\partial\boldsymbol{F}^{\mathrm{cf}}/\partial M\) cannot be expressed as a linear combination of the columns of \(\partial\boldsymbol{F}^{\mathrm{cf}}/\partial\boldsymbol{y}^{\mathrm{cf}}\)~\cite{Calvetti2000}, 
augmenting Eq.~\eqref{eq:F_param} with one scalar equation that supplies information along this missing direction is sufficient to restore full rank.

To this end, the additional equation is constructed starting from
\begin{equation}
\label{eq:eq_normalization_differential}
\left.
\left(
\frac{\mathrm{d}\hat{\boldsymbol{y}}^{\mathrm{cf}}}{\mathrm{d}s}
\right)^{\!\top}
\right|_{s_0}
\frac{\mathrm{d}\hat{\boldsymbol{y}}^{\mathrm{cf}}}{\mathrm{d}s}
=
1\,,
\end{equation}
which is the differential form of the arc‑length constraint where \(s_0\) denotes the current point on the solution curve. Integrating from \( s_0 \) to the next solution point \( s^* \) gives the pseudo‑arc‑length normalization condition
\begin{equation}
\label{eq:eq_normalization}
N =
\left.
\left(
\frac{\mathrm{d}\hat{\boldsymbol{y}}^{\mathrm{cf}}}{\mathrm{d}s}
\right)^{\!\top}
\right|_{s_0}
\!\!
\left(
\hat{\boldsymbol{y}}^{\mathrm{cf}}(y,s^*)
-
\hat{\boldsymbol{y}}^{\mathrm{cf}}_0(y)
\right)
-
\Delta s
=
0\,,
\end{equation}
where \(\hat{\boldsymbol{y}}^{\mathrm{cf}}_0(y)=\hat{\boldsymbol{y}}^{\mathrm{cf}}(y,s_0)\) is the solution at \(s_0\), and \( \Delta s\) is the prescribed step size. 

Equations~\eqref{eq:F_param} and~\eqref{eq:eq_normalization} together define a closed augmented boundary value problem, written compactly as
\begin{equation}
\label{eq:cfbvp_recast}
\hat{\boldsymbol{F}}^{\mathrm{cf}}\!\left(\hat{\boldsymbol{y}}^{\mathrm{cf}}(y,s),\,
\hat{\boldsymbol{y}}^{\mathrm{cf}}_0(y),\,
\left.\frac{\mathrm{d}\hat{\boldsymbol{y}}^{\mathrm{cf}}}{\mathrm{d}s}\right|_{s_0},\,
\boldsymbol{y}^{\mathrm{cf}}_{\mathrm{l}},\,
\boldsymbol{y}^{\mathrm{cf}}_{\mathrm{r}},\,
\Delta s,\,
\boldsymbol{\alpha}\right)=0\,.
\end{equation}
The Jacobian of the augmented system~\eqref{eq:cfbvp_recast} is non-singular at the turning point. 
The row of the augmented Jacobian corresponding to the pseudo-arc-length normalization condition is, by construction, equal to
\(\left(\mathrm{d}\hat{\boldsymbol{y}}^{\mathrm{cf}}/\mathrm{d}s\right)^{\!\top}|_{s_0}\), so its inner product with the tangent vector is
\begin{equation}
\left.
\left(\frac{\mathrm{d}\hat{\boldsymbol{y}}^{\mathrm{cf}}}{\mathrm{d}s}\right)^{\!\top}
\right|_{s_0}
\frac{\mathrm{d}\hat{\boldsymbol{y}}^{\mathrm{cf}}}{\mathrm{d}s}
=
\left\|\frac{\mathrm{d}\boldsymbol{y}^{\mathrm{cf}}}{\mathrm{d}s}\right\|^2
=
1\,,
\end{equation}
where the first equality follows from \(\mathrm{d}\hat{\boldsymbol{y}}^{\mathrm{cf}}/\mathrm{d}s = (\mathrm{d}\boldsymbol{y}^{\mathrm{cf}}/\mathrm{d}s,\, 0)^{\top}\) at the turning point. Since this inner product is nonzero, the additional row does not belong to the row space of \(\partial\boldsymbol{F}^{\mathrm{cf}}/\partial\boldsymbol{y}^{\mathrm{cf}}\) and thus supplies the missing row. Together with the earlier condition that \(\partial\boldsymbol{F}^{\mathrm{cf}}/\partial M\) does not lie in the range of \(\partial\boldsymbol{F}^{\mathrm{cf}}/\partial\boldsymbol{y}^{\mathrm{cf}}\), which supplies the missing column, the augmented Jacobian is of full rank.

In the present implementation, the normalization condition is written in terms of the reduced quantity \(\boldsymbol q\left(\hat{\boldsymbol y}^{\mathrm{cf}}\right) = \left(\ln (M), T_{\max}\right)^{\top}\) instead of the full augmented solution vector. The maximum temperature is defined by \(T_{\max}(s)=\max_{y} T(y,s)\). Since \(M>0\), \(\ln (M)\) can be used instead of \(M\) without changing the turning point. %
Further implementation details can be found in the continuation formulations of Giovangigli and Smooke~\cite{Giovangigli1989, Giovangigli1993, Giovangigli1987Onestep}.

\subsection{Sensitivity analysis}

In this work, sensitivity coefficients are computed using the adjoint discrete formulation~\cite{Langer2021}, which requires substantially less
implementation effort than the continuous approach. In this formulation,
the adjoint equations are derived directly from the discretized closed
homogeneous reactor and counterflow flame systems~(Eqs.~\eqref{eq:hrbvp_recast} and~\eqref{eq:cfbvp_recast}), which can be
expressed compactly as
\begin{equation}
\label{eq:compact}
\hat{\boldsymbol{F}}^{\kappa}\left(\hat{\boldsymbol{y}}^{\kappa}, \boldsymbol{\alpha}\right) = 0,
\qquad
\kappa \in \left\{\mathrm{hr},\mathrm{cf}\right\}.
\end{equation}
The superscript \(\kappa\) and functional dependencies are omitted below whenever the same statement applies to both systems.

Given a scalar objective function 
\begin{equation} \label{eq:objective}
J = J(\hat{\boldsymbol{y}}, \boldsymbol{\alpha}),
\end{equation}
which expresses the explicit dependence of the QoI on \(\hat{\boldsymbol{y}}\) and the input parameters \(\boldsymbol{\alpha}\), the logarithmic sensitivity coefficient of \(J\) with respect to the input parameter \(\alpha_i\) is defined as
\begin{equation} 
S_{\mathrm{QoI},\alpha_i}= \frac{\alpha_i}{J} \frac{\mathrm{d}J}{\mathrm{d}\alpha_i}.
\end{equation}

The derivation of \(\mathrm{d}J\text{/}\mathrm{d}\alpha_i\) starts from the total differentials of Eqs.~\eqref{eq:compact} and~\eqref{eq:objective}, given as
\begin{equation} 
\label{eq:dL}
\mathrm{d} \hat{\boldsymbol F}(\hat{\boldsymbol y}, \boldsymbol{\alpha})  =\frac{\partial \hat{\boldsymbol F}}{\partial \hat{\boldsymbol y}} \mathrm{d}\hat{\boldsymbol y}+\frac{\partial \hat{\boldsymbol F}}{\partial \boldsymbol{\alpha}} \mathrm{d} \boldsymbol{\alpha}=0
\end{equation}
and
\begin{equation} \label{eq:dJ}
\mathrm{d} J(\hat{\boldsymbol y}, \boldsymbol{\alpha})  =\frac{\partial J}{\partial \hat{\boldsymbol y}} \mathrm{d} \hat{\boldsymbol y}+\frac{\partial J}{\partial \boldsymbol{\alpha}} \mathrm{d} \boldsymbol{\alpha}\, . 
\end{equation}
In the adjoint approach, following Eq.~\eqref{eq:dL}, an auxiliary variable \(\boldsymbol{\varphi}\) can be introduced, such that

\begin{align}
\begin{split} 
\label{eq:dJ_intermediate}
\mathrm{d} J(\hat{\boldsymbol{y}}, \boldsymbol{\alpha}) 
&=
\frac{\partial J}{\partial \hat{\boldsymbol{y}}} \, \mathrm{d} \hat{\boldsymbol{y}}
+ \frac{\partial J}{\partial \boldsymbol{\alpha}} \, \mathrm{d} \boldsymbol{\alpha}
- \boldsymbol{\varphi}^{\top}
\left(
\frac{\partial \hat{\boldsymbol{F}}}{\partial \hat{\boldsymbol{y}}} \, \mathrm{d} \hat{\boldsymbol{y}}
+ \frac{\partial \hat{\boldsymbol{F}}}{\partial \boldsymbol{\alpha}} \, \mathrm{d} \boldsymbol{\alpha}
\right) \\
&= \left(\frac{\partial J}{\partial \hat{\boldsymbol{y}}}-\boldsymbol{\varphi}^{\top} \frac{\partial \hat{\boldsymbol F}}{\partial \hat{\boldsymbol{y}}}\right) \mathrm{d} \hat{\boldsymbol{y}}+\left(\frac{\partial J}{\partial \boldsymbol{\alpha}}-\boldsymbol{\varphi}^{\top} \frac{\partial \hat{\boldsymbol F}}{\partial \boldsymbol{\alpha}}\right) \mathrm{d} \boldsymbol{\alpha}.
\end{split} 
\end{align}
Choosing $\boldsymbol{\varphi}$ to satisfy 
\begin{equation}
\label{eq:syst_adj}
\left[\frac{\partial \hat{\boldsymbol F}}{\partial \hat{\boldsymbol{y}}}\right]^{\top} \boldsymbol{\varphi}=\left(\frac{\partial J}{\partial \hat{\boldsymbol{y}}}\right)^{\top} ,
\end{equation}
the dependence in Eq.~\eqref{eq:dJ_intermediate} of \(\mathrm{d}J\) on \(\hat{\boldsymbol y}\) is eliminated, and \(\mathrm{d} J \text{/} \mathrm{d} \alpha_i\) can be obtained as 
\begin{equation}
\label{eq:adj}
\frac{\mathrm{d} J}{\mathrm{d} \alpha_i}=\frac{\partial J}{\partial \alpha_i}-\boldsymbol{\varphi}^{\top} \frac{\partial \hat{\boldsymbol F}}{\partial \alpha_i}.
\end{equation}
The sensitivity coefficients in the forward approach~\cite{Reuven1986} are computed from Eq.~\eqref{eq:dJ} as 
\begin{equation} \label{eq:forw}
\frac{\mathrm{d} J}{\mathrm{d} \alpha_i}=\frac{\partial J}{\partial \hat{\boldsymbol y}} \frac{\mathrm{d}\hat{\boldsymbol y}}{\mathrm{d} \alpha_i}+\frac{\partial J}{\partial \alpha_i}
\end{equation} 
after solving 
\begin{equation}
\label{eq:syst_forw}
\left[\frac{\partial \hat{\boldsymbol F}}{\partial \hat{\boldsymbol y}}\right] \frac{\mathrm{d} \hat{\boldsymbol y}}{\mathrm{d} \alpha_i} = - \frac{\partial \hat{\boldsymbol F}}{\partial \alpha_i},  
\end{equation}
which is obtained from Eq.~\eqref{eq:dL}.

\section{Computational time-complexity analysis}
\label{section:ctca}

The linear systems in Eqs.~\eqref{eq:syst_adj} in the adjoint approach and~\eqref{eq:syst_forw} in the forward approach are solved using a direct method based on LU factorization of the Jacobian matrix \( \left[\mathrm{d}\hat{\boldsymbol F}/\mathrm{d}\hat{\boldsymbol y}\right] \), followed by forward and backward substitutions to solve the resulting triangular systems. In the adjoint approach, sensitivity coefficients of QoI with respect to the model parameters are obtained by solving the linear system in Eq.~\eqref{eq:syst_adj} once for each QoI. In contrast, the forward approach requires one linear solve per model parameter, since the right-hand side of Eq.~\eqref{eq:syst_forw} depends on the parameter being considered. This difference underlines the superior algorithmic efficiency of the adjoint approach, as discussed in more detail below.

In both the forward and the adjoint approaches, the dominant computational cost arises from three main operations:
\begin{enumerate}
    \item numerical evaluation of the Jacobian matrix,
    \item LU factorization of the Jacobian matrix, 
    \item forward-backward substitution. 
\end{enumerate}

Exploiting the banded structure of the Jacobian matrix, the asymptotic theoretical computational cost of the adjoint approach is 
\begin{equation}
\label{eq:cost_adj}
C_{\mathrm{adj}} \sim \mathcal{O}\!\left(n_{\mathrm{gp}}\, n_{\mathrm{eq}}^{2}\right) + 
\mathcal{O}\!\left(n_{\mathrm{gp}}\, n_{\mathrm{eq}}^{3}\right)
+ \mathcal{O}\!\left(n_{\mathrm J}\, n_{\mathrm{gp}}\, n_{\mathrm{eq}}^{2}\right)\, ,
\end{equation}
where \( n_{\mathrm{gp}} \) denotes the number of grid points used for the discretization and \( n_{\mathrm{eq}} \) the number of dependent variables (cf. Eqs.~\eqref{eq:y_hr_hat} and~\eqref{eq:y_cf_hat}). Here, the first term corresponds to the Jacobian evaluation, the second to its LU factorization, and the third to the cumulative cost of the forward-backward substitutions, which must be performed \(n_{\mathrm J}\) times, where \(n_{\mathrm J}\) denotes the number of objective functions.
Similarly, for the forward approach,
\begin{equation}
\label{eq:cost_forw}
C_{\mathrm{forw}} \sim  \mathcal{O}\!\left(n_{\mathrm{gp}}\, n_{\mathrm{eq}}^{2}\right)+
\mathcal{O}\!\left(n_{\mathrm{gp}}\, n_{\mathrm{eq}}^{3}\right)
+ \mathcal{O}\!\left(n_{\mathrm p}\, n_{\mathrm{gp}}\, n_{\mathrm{eq}}^{2}\right),
\end{equation}
since the forward-backward substitution must be performed \(n_{\mathrm p}\) times, where \(n_{\mathrm p}\) denotes the number of parameters.

Because \(n_{\mathrm J}\) is typically much smaller than \(n_{\mathrm{eq}}\), Eq.~\eqref{eq:cost_adj} shows that the LU factorization term governs the asymptotic computational complexity of the adjoint approach. Another important observation is that the numerical algorithms commonly used to compute the primary solutions for the configurations considered here rely on Newton's method and therefore require factorizing the same Jacobian matrix as in the sensitivity analysis. Consequently, the dominant asymptotic cost of the adjoint sensitivity analysis presented here is not greater than that of computing the primary solution itself.

No analogous guarantee exists for the forward approach, since Eq.~\eqref{eq:cost_forw} retains an explicit dependence on \(n_{\mathrm{p}}\). In combustion applications, \(n_{\mathrm{p}}\) scales at least linearly with the model size, and hence with \(n_{\mathrm{eq}}\), because thermodynamic data, reaction-rate parameters, and most transport inputs contribute a fixed number of parameters per species or reaction. With \(n_{\mathrm{p}} > n_{\mathrm{eq}}\), the forward-backward substitution dominates the asymptotic cost of the forward approach, and the theoretical speedup of the adjoint method follows as
\begin{equation}
\label{eq:adj_speedup}
C_{\mathrm{su}} \sim \mathcal{O}\!\left(\frac{n_{\mathrm{gp}}\, n_{\mathrm{p}}\, n_{\mathrm{eq}}^{2}}{n_{\mathrm{gp}}\, n_{\mathrm{eq}}^{3}}\right)
= \mathcal{O}\!\left(\frac{n_{\mathrm{p}}}{n_{\mathrm{eq}}}\right).
\end{equation}

For the IDT and ESR problems considered here \(n_{\mathrm{J}} = 1\); therefore, the reformulated adjoint approaches are never asymptotically worse than the conventional forward approach and are often considerably faster. In practice, however, the cost is strongly affected by implementation details beyond the reach of asymptotic analysis; these are examined in the next section.

Compared with the brute-force approach, both the forward and adjoint methods avoid repeating simulations per parameter and thus reduce the number of Jacobian factorizations by a factor of \(n_{\mathrm p}\). The substantial speedup offered by forward methods over the brute-force approach has been extensively analyzed in the literature~\cite{Lemke2019,Almohammadi2021,dsa_2019}; therefore, only the relative advantages of the adjoint method over the forward method are discussed below.

\section{Implementation details}

The algorithms for computing both forward and adjoint sensitivity coefficients are implemented in FlameMaster~\cite{FlameMaster}, a C++ code used to simulate zero-dimensional reactors and one-dimensional flames.

For the IDT problem, linear systems are solved with the \textsc{KINSOL} solver of the \textsc{SUNDIALS} suite~\cite{hindmarsh2005sundials}, which relies on \textsc{LAPACK}~\cite{LAPACK}
routines linked to Intel \textsc{MKL} (2024.2.0) for banded matrix factorization and triangular solves.
The derivatives required for the computation of the sensitivity coefficients in Eqs.~\eqref{eq:adj} and~\eqref{eq:forw} are evaluated through a hybrid analytical-numerical approach. 
In the adjoint implementation, instead of forming and storing the full matrix \(\partial \hat{\boldsymbol{F}}/\partial \boldsymbol{\alpha}\), a single sweep over the grid evaluates, at each grid point \(j\), the dot product between the adjoint auxiliary variable block \(\boldsymbol{\varphi}_{j}\) and the corresponding \(\partial \hat{\boldsymbol{F}}_{j}/\partial \boldsymbol{\alpha}\) contribution.

For the ESR problem, LU factorization and forward-backward substitutions are performed using an in-house implementation that exploits the block-tridiagonal structure~\cite{Hindmarsh1977} of the Jacobian matrix arising from the discretized system in Eq.~\eqref{eq:cfbvp_recast}. 
The implementation uses the Eigen~\cite{eigen} library to handle the small dense Jacobian blocks, while block-level LU factorizations and triangular solves are performed using \textsc{BLAS}/\textsc{LAPACK}~\cite{LAPACK} routines as implemented by Intel's \textsc{MKL} (2024.2.0). 
The derivatives in Eqs.~\eqref{eq:adj} and~\eqref{eq:forw} are computed via algorithmic differentiation using the dco/c++ library~\cite{Naumann2012, Leppkes2016}. In particular, the adjoint mode of dco/c++ allows the product \(\boldsymbol{\varphi}^{\top}\left[\partial \hat{\boldsymbol F}/\partial \boldsymbol{\alpha}\right]\) in Eq.~\eqref{eq:adj} to be computed implicitly, without assembling the full \(\partial \hat{\boldsymbol F}/\partial \boldsymbol{\alpha}\) matrix. 

\section{Results and discussion}
The adjoint sensitivity coefficients for both the ignition delay time and extinction strain rate are first compared against those obtained using the forward and brute-force approaches to verify the consistency of the three different methods.

Execution time measurements then quantify the practical performance gains of the adjoint approach over the forward method. Finally, an operation-level performance analysis of the IDT problem compares the current implementation against the theoretical complexity estimates, assessing the extent to which the predicted algorithmic efficiency is realized in practice.
The analysis identifies the major bottlenecks and highlights the operations that should be prioritized in future efficiency improvement efforts.

\subsection{Comparison of adjoint, forward, and brute-force sensitivity coefficients}

In the brute-force approach, first-order local sensitivity coefficients are approximated using the logarithmic finite-difference formulation
\begin{equation}
\label{eq:bf_idt}
    S_{\mathrm{QoI},\alpha_i}^{\mathrm{BF}}
    =
    \frac{\alpha_i}{Q_0}\cdot
    \frac{\Delta Q}{\Delta \alpha_i}
    =
    \frac{\alpha_i}{Q_0}\cdot
    \frac{\Delta Q}{\alpha_i(\gamma-1)},
\end{equation}
where \(Q_0\) is the quantity of interest evaluated from the reference simulation, \(\Delta Q\) is the difference between the quantities of interest evaluated from the perturbed and the reference simulation, and \(\gamma\) is the perturbation factor.
For both QoIs, the comparison is restricted to sensitivity coefficients with respect to the frequency factors of Arrhenius-type reactions and the high-pressure limits of pressure-dependent fall-off reactions, to reduce the computational cost of the brute-force analysis.

Agreement between the different methods is assessed qualitatively through parity diagrams of the sensitivity coefficients and quantitatively using the cosine similarity,
\begin{equation}
    \epsilon =
    \frac{\mathbf{s}_1\cdot\mathbf{s}_2}
    {\left\|\mathbf{s}_1\right\|_2
    \left\|\mathbf{s}_2\right\|_2},
\end{equation}
which measures the alignment between the sensitivity coefficient vectors \(\mathbf{s}_1\) and \(\mathbf{s}_2\) obtained from the different methods~\cite{Ji20202}. Here \(\left\|\cdot\right\|_2\) denotes the Euclidean norm, and the \(i\)-th component of a sensitivity coefficient vector corresponds to the sensitivity coefficient with respect to parameter \(\alpha_i\).

In contrast to direct comparison of individual coefficient magnitudes, the cosine similarity quantifies the consistency with which the different methods identify and rank the dominant kinetic parameters. This criterion is particularly appropriate here because differences in coefficient values are expected owing to the fundamentally different formulations and numerical procedures of the direct methods compared to the brute-force approach.

The IDT is defined implicitly from the solution vector as the time at which a user-defined ignition criterion is satisfied, e.g., the time at which the derivative of the temperature reaches its maximum. Four different definitions of IDT are considered, and for each of them, the temperature at the corresponding ignition point is imposed as a boundary condition for the augmented system in Eq.~\eqref{eq:ivp_recast}.
The ignition delay time obtained from the primal IVP solution, together with the corresponding solution trajectory from \(t_0 = 0\) to \(\tau\), is then used to construct the initial guess for the Newton iteration, which is used to solve the BVP.

The objective function adopted for the forward and adjoint methods is 
defined as
\begin{equation}
\label{eq:j_idt}
  J(\hat{\boldsymbol y}^{\mathrm{hr}},\boldsymbol{\alpha})
  = \tau.
\end{equation}
 Upon discretization, \(\tau\) appears as a component \(\tau_\mathrm{idx}\) of the solution vector, one per grid point \(\mathrm{idx}\). Since it is an eigenvalue of the system, its value is the same across all grid points. Consequently, \(\tau\) can be evaluated at any grid point, without affecting the sensitivity coefficient values~\cite{dsa_2019}.

For the brute-force method, a single global perturbation factor was used for all the considered \(\alpha_i\). Its value was chosen to lie within the range where the perturbation is large enough to avoid floating-point errors, yet small enough to remain in the linear-response regime.
To this end, sensitivity coefficients for each \(\alpha_i\) were evaluated over a range of perturbation factors, and plateau regions were identified where the coefficients remained approximately constant. The global perturbation factor was then determined as a weighted average of the centers of these plateau regions. Larger weights were assigned to
reactions exhibiting both high sensitivity coefficients and well-defined plateau regions, while oscillatory sensitivities, typically associated with low-magnitude coefficients and numerical noise, received less weight. A detailed description of the procedure used to determine the perturbation factor employed for the brute-force method is provided in the Supplementary Material~(SM).

\begin{figure*}[!htbp]
    \centering
     \includegraphics[width=\textwidth]{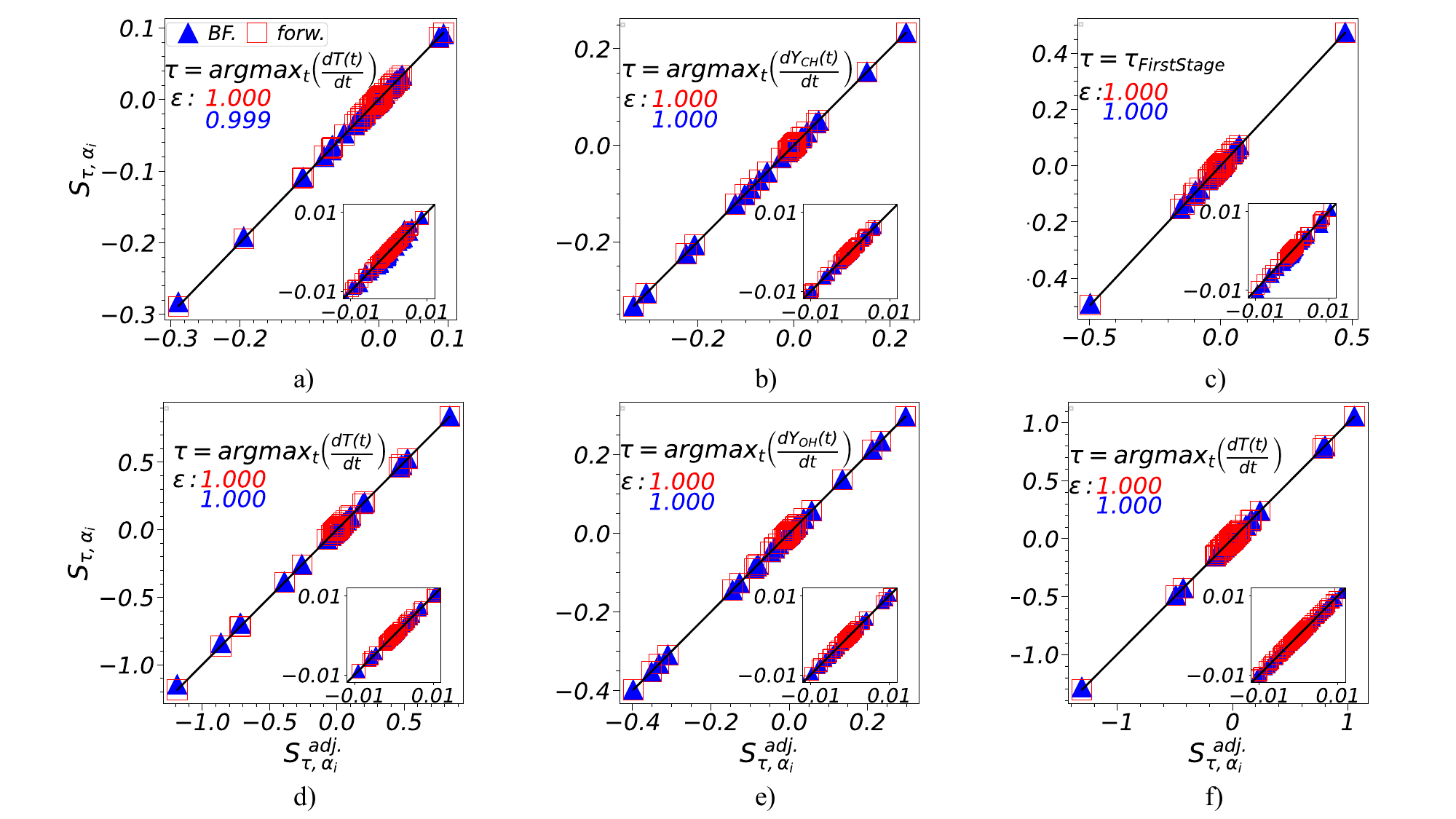}
  \caption{Parity diagrams comparing sensitivity coefficients computed using adjoint, forward~(red), and brute-force~(blue) methods for different IDT definitions. Conditions are:~(a) \ce{NH3}/air, \(T=\SI{2000}{\kelvin}\), \(\phi=1\), \(p=\SI{1}{\bar}\), \(\gamma = 1.04 \), kinetic model from Ref.~\cite{Zhu2024};~(b) \ce{CH4}/air, \(T=\SI{1250}{\kelvin}\), \(\phi=1.5\), \(p=\SI{1}{\bar}\), \(\gamma = 1.003 \), kinetic model from ~\cite{c31,c32}~(C2 module of C3MechV4.0.1);~(c) \ce{C3H8}/air, \(T=\SI{650}{\kelvin}\), \(\phi=1\), \(p=\SI{30}{\bar}\), \(\gamma = 1.033\), kinetic model from Ref.~\cite{Murakami2025}. First-stage ignition is identified as the time of maximum temperature-rise rate ($\mathrm{d}T/\mathrm{d}t$) occurring before the mixture has heated \SI{250}{\kelvin} above its initial value;~(d) \ce{NH3}/air, \(T=\SI{1100}{\kelvin}\), \(\phi=0.6\), \(p=\SI{10}{\bar}\), \(\gamma = 1.03\), kinetic model from Ref.~\cite{Zhu2024};~(e) \ce{CH4}/air, \(T=\SI{1250}{\kelvin}\), \(\phi=0.5\), \(p=\SI{1}{\bar}\), \(\gamma = 1.002 \);~(f) \ce{C3H8}/\ce{O2}/\ce{Ar}/\ce{N2}/\ce{NO2}, \(T=\SI{769}{\kelvin}\), \(\phi=1\), \(p=\SI{30}{\bar}\), \(\gamma = 1.011 \), kinetic model from Ref.~\cite{Murakami2025}.}
  \label{fig:idt_validation}
\end{figure*}

Sensitivity coefficients obtained for several cases representative of different combustion regimes are compared in Fig.~\ref{fig:idt_validation}, together with the corresponding cosine similarity. Two cosine similarities are reported for each case: one comparing the adjoint and the forward sensitivity coefficient vectors~(red), and the other comparing the adjoint and the brute-force sensitivity vectors~(blue).
Across all cases, \(\epsilon\) remains close to unity, indicating 
that all three sensitivity methods are consistent in identifying the dominant parameters controlling the auto-ignition process. Furthermore, the agreement across the methods for the different definitions of \(\tau\) confirms that the proposed adjoint formulation retains the flexibility of the forward sensitivity analysis approach~\cite{dsa_2019}, yielding accurate sensitivity coefficients regardless of the specific ignition-delay definition adopted.

For the extinction strain rate problem, the QoI is defined as the strain rate value at the turning point. The turning point is identified by a sign change in the variation of \(\ln M\) between two consecutive arc-length steps.

For the adjoint and forward methods, the objective function is:
\begin{equation}
\label{eq:j_esr}
  J(\hat{\boldsymbol y}^{\mathrm{cf}},\boldsymbol{\alpha})
  =
  \frac{4}{L}
  \sqrt{\frac{M(s_{\mathrm{ext}})}
  {\rho_{\mathrm{r}}}},
\end{equation}
where \(s_{\mathrm{ext}}\) denotes the arclength coordinate of the last converged solution reached before the turning point, as the solution branch is continued from the stable branch toward the extinction fold. The quantity \(M\) is recovered from the variable \(\ln M\), which augments the solution vector of
the recast system in Eq.~\eqref{eq:cfbvp_recast}. After discretization, \(\ln M\) appears
as an entry, \(\ln M_{\mathrm{idx}}\), in the solution vector at each grid point
\(\mathrm{idx}\). The arclength-continuation normalization condition in Eq.~\eqref{eq:eq_normalization} is enforced
at one grid point~(i.e., the grid point where the flame temperature reaches its maximum),  whereas the constraint
\(\mathrm{d}\ln M/\mathrm{d}y = 0\) is imposed at all other grid points to preserve
the block-tridiagonal structure of the Jacobian matrix. Consequently, \(\ln M\)
remains constant throughout the grid. It may therefore be evaluated at any grid
point, with no effect on the results. 

\begin{figure}[htbp]
\centering
    \includegraphics[trim={40pt 0pt 0pt 0pt}, clip, width=165pt]{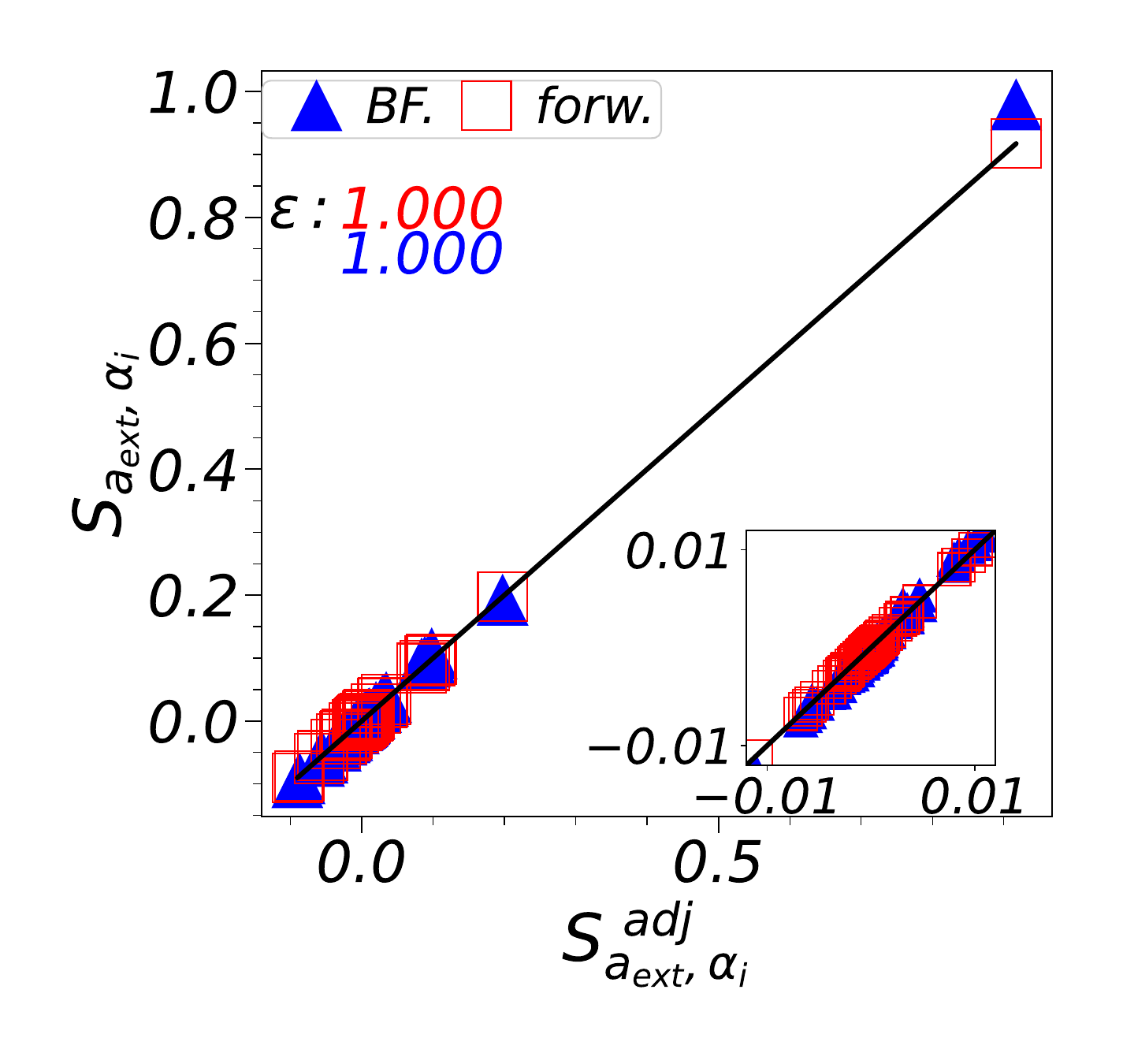}
    \caption{Parity diagram of sensitivity coefficients for the ESR computed using the adjoint, forward~(red), and brute-force~(blue) methods. Results are obtained for a \ce{CH4}/air counterflow flame at \(p=\SI{1}{\bar}\).
   Left side: \(T_{\mathrm{l}}=\SI{295}{\kelvin}\), \(v_{\mathrm{l}}=\SI{0.6}{\meter\per\second}\), mass flux fractions \ce{CH4} 0.0360 / \ce{CO} 0.0161 / \ce{H2} 0.0085 / \ce{CO2} 0.0169 / \ce{N2} 0.9225, \( L = \SI{1}{\centi\meter}\).
   Right side: \(T_{\mathrm{r}}=\SI{295}{\kelvin}\), \(v_{\mathrm{r}}=\SI{-0.54}{\meter\per\second}\), mass flux fractions \ce{O2} 0.2330 / \ce{N2} 0.7670. The brute-force method uses \(\gamma=1.05\). The kinetic model is the C2 module of C3MechV4.0.1~\cite{c31,c32}.}
\label{fig:esr_validation}
\end{figure}

A comparison between the sensitivity coefficients obtained from the three different methods for extinction strain rate is depicted in Fig.~\ref{fig:esr_validation}.
Overall, satisfactory agreement is observed across all methods, with the forward and adjoint methods once again showing closer mutual agreement than with the brute-force approach, as they are mathematically identical.

For the ESR problem, the dependence of the brute-force sensitivity coefficients on the perturbation factor is more pronounced than for the IDT problem. Depending on the perturbation factor value, the perturbed extinction curves may deviate significantly from the reference curve. This change in boundary conditions could account for the even more pronounced difference between the adjoint and the brute-force methods compared to that between the adjoint and the forward methods, since forward and adjoint sensitivities are instead evaluated numerically under the same boundary conditions.

\subsection{Performance evaluation}

Benchmarks were conducted in single-threaded mode on the
CLAIX-2023-HPC system at RWTH Aachen University, equipped with two Intel Xeon Platinum~8468 Sapphire Rapids processors (96 cores in total) and 256~GB of main memory.
The code was compiled with the Intel oneAPI toolchain using performance-oriented flags~(\texttt{-O3}, \texttt{-ffast-math}, \texttt{-DNDEBUG}). 

For the IDT problem, total wall-clock times for determining the sensitivity coefficients were measured for a stoichiometric~\ce{CH4}/\ce{air} mixture at \SI{1250}{K} and atmospheric pressure, simulated using kinetic models of increasing size extracted from C3MechV4.0.1~\cite{c31,c32}.
To emphasize the effect of parameter-dependent scaling of the forward over the adjoint method, all reversible reactions were treated as independent forward and backward reactions. The backward Arrhenius parameters were obtained by fitting reverse rate coefficients computed during preprocessing from the forward rate and the corresponding equilibrium constants. 
The measured wall-clock times for the forward and adjoint methods are reported in Tab.~\ref{tab:cpu-comparison}.

\begin{table}[h!]
\caption{Wall-clock times for computing \(S_{\tau,\alpha_i}\) with the forward and adjoint methods for \(n_{\mathrm p}\) parameters (\ce{CH4}/\ce{air}, \SI{1250}{K}, \(\phi = 1\), \SI{1}{atm}). The mechanism identifier (MID) refers to the specific {C3MechV4.0}~\cite{c31,c32} submodule used.
Labels with HT denote kinetic models that include only high-temperature chemistry, while N indicates the inclusion of the \ce{N}- and N-C-chemistry submodule.
The speedup~(\(\mathcal{S}\)) is defined as the ratio between the adjoint and forward wall-clock time.}
\centering
\footnotesize
\setlength{\tabcolsep}{2pt}
\begin{tabular}{ccccccc}
\hline
\textbf{Mech.} & MID & \textbf{$n_{\mathrm p}$} & \textbf{$n_{\mathrm{eq}}$} &
\textbf{Time} & \textbf{Time} &
\textbf{\(\mathcal{S}\)} \\
          &  &      &     & \textbf{adj.} \textbf{[\SI{}{s}]}  & \textbf{forw.} \textbf{[\SI{}{s}]}       & \\
\hline\hline
C4-HT & 3R00   & 5000   &380  & 8.24  & 521    & 63.2 \\
C7-HT-N & 3W28 & 11623  &1027 & 61.8 & 8730   & 141  \\
C6     & 4EI0  & 17344  &2301 & 358   & 65700  & 183  \\
C7     & 4EM0  & 19661  &2743 & 643   & 136000 & 211  \\
C8     &  4ENC & 30320  &4637 & 925   & 258000 & 279  \\
\hline
\end{tabular}
\label{tab:cpu-comparison}
\end{table}

For the ESR problem, total wall-clock times were measured for the \ce{CH4}/\ce{air} counterflow configuration used for method verification (Fig.~\ref{fig:esr_validation}), again using kinetic submodels of increasing size derived from C3MechV4.0.1~\cite{c31,c32}.
Sensitivities were computed with respect to the forward frequency factors.
The corresponding wall-clock times for the forward and adjoint methods are reported in Tab.~\ref{tab:cpu-comparison-esr}.

\begin{table}[h!]
\caption{Wall-clock times for computing \(S_{a_{\mathrm{ext}},\alpha_i}\) with the forward and adjoint methods for \(n_{\mathrm p}\) parameters, for the counterflow configuration of Fig.~\ref{fig:esr_validation}.The mechanism identifier (MID) refers to the specific {C3MechV4.0}~\cite{c31,c32} submodule used.
Labels ending with HT denote kinetic models including only high-temperature chemistry.
The speedup~(\textbf{\(\mathcal{S}\)}) is defined as in Tab.~\ref{tab:cpu-comparison}.}
\centering
\footnotesize
\setlength{\tabcolsep}{2pt}
\begin{tabular}{ccccccc}
\hline
\textbf{Mech} & MID & \textbf{$n_{\mathrm p}$} & \textbf{$n_{\mathrm{eq}}$} &
\textbf{Time adj.} & \textbf{Time forw.} &
\textbf{\(\mathcal{S}\)} \\
         &   &    &      &\textbf{[\SI{}{s}]}  & \textbf{[\SI{}{s}]}       & \\
\hline\hline
C4-HT   & 3R00 & 1928 &382  & 26.5 & 542  & 20.5 \\
C4      & 44I0 & 3126 &733  & 99.0 & 2110 & 21.3 \\
C6-HT   & 3W00 & 3605 &823& 117  & 2930 & 25.0 \\
C8-HT   & 3W2O & 6157 &1411 & 395 & 12900 & 32.7 \\
\hline
\end{tabular}
\label{tab:cpu-comparison-esr}
\end{table}

Tables~\ref{tab:cpu-comparison} and~\ref{tab:cpu-comparison-esr} show that the adjoint speedups range from one to more than two orders of magnitude relative to the forward method.
For the largest considered model in the IDT case (C8, 30320 parameters), the forward method requires about ~\SI{71}{h} of wall-clock time compared with only~\SI{15}{min} for the adjoint method. 
This substantial difference is operationally relevant for workflows that utilize sensitivity analysis for large kinetic models, such as mechanism reduction, optimization, design of experiments, or uncertainty quantification.

Despite these substantial speedups, the wall-clock times in Tabs.~\ref{tab:cpu-comparison} and~\ref{tab:cpu-comparison-esr} capture not only algorithmic complexity but also implementation and hardware-dependent effects. To assess how the latter limit the realization of the theoretical efficiency gains in practice, the measured wall-clock times of the major computational operations are compared in the following section with the scaling laws derived in Sec.~\ref{section:ctca}. This analysis identifies the operations that dominate the overall runtime and thereby highlights the most promising areas for further optimization. For brevity, this performance analysis is restricted to the IDT problem. The operation breakdown and scaling laws for the ESR problem are briefly discussed in the SM.

\subsubsection{Adjoint performance analysis}

\begin{figure}[!h]
\centering
\includegraphics[width=\linewidth]{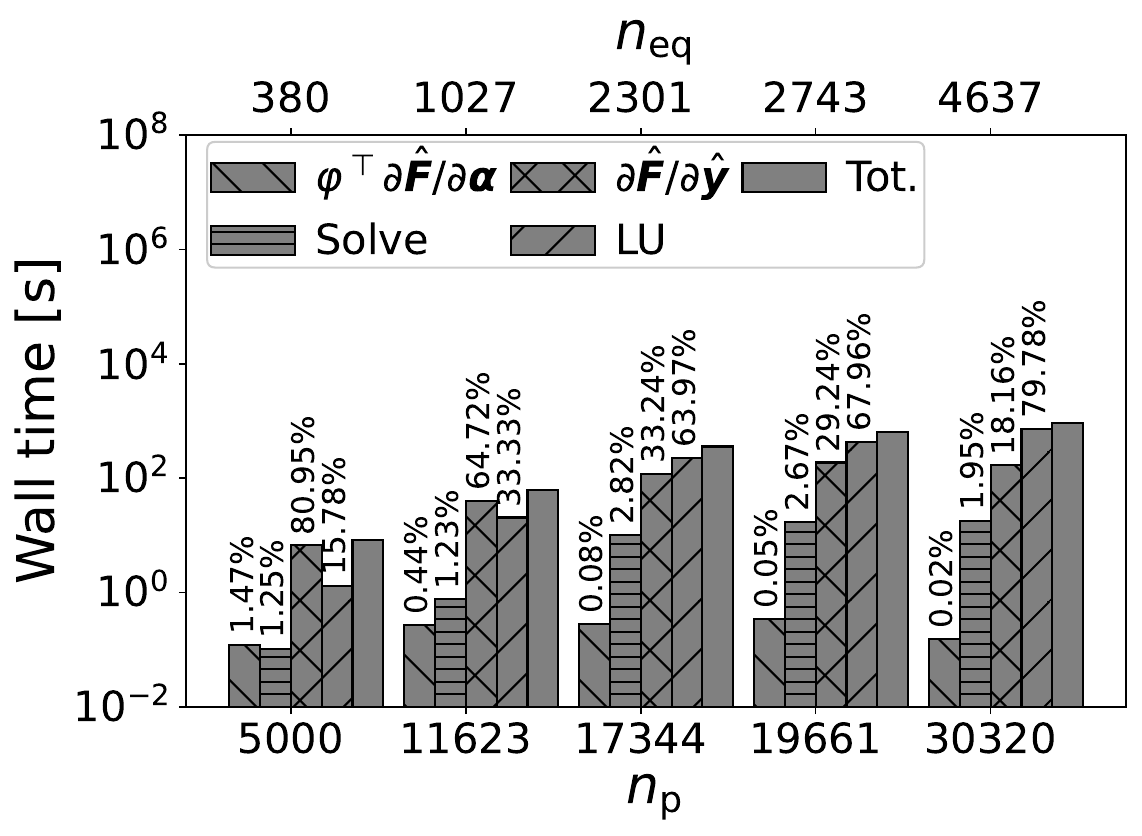}
\caption{Relative contribution of the main computational operations to the total adjoint computational cost for each benchmarked kinetic model.}
\label{fig:idt_adj_contribution}
\end{figure}

Figure~\ref{fig:idt_adj_contribution} shows the relative contribution of the dominant operations to the total wall-clock time for the adjoint sensitivity computation with each considered kinetic model. For the smallest model, Jacobian evaluation accounts for~81\% of the total runtime. As the problem size increases, LU factorization becomes increasingly dominant, contributing about 80\% for the largest model, in agreement with the analysis in Sec.~\ref{section:ctca}. All remaining operations contribute less than~3\% for all the models.
The overall adjoint cost is therefore governed by these two operations, and the extent to which the theoretical asymptotic behavior is realized depends on the problem size at which LU factorization overtakes Jacobian evaluation.

\begin{figure}[!h]
\centering
\includegraphics[width=\linewidth]{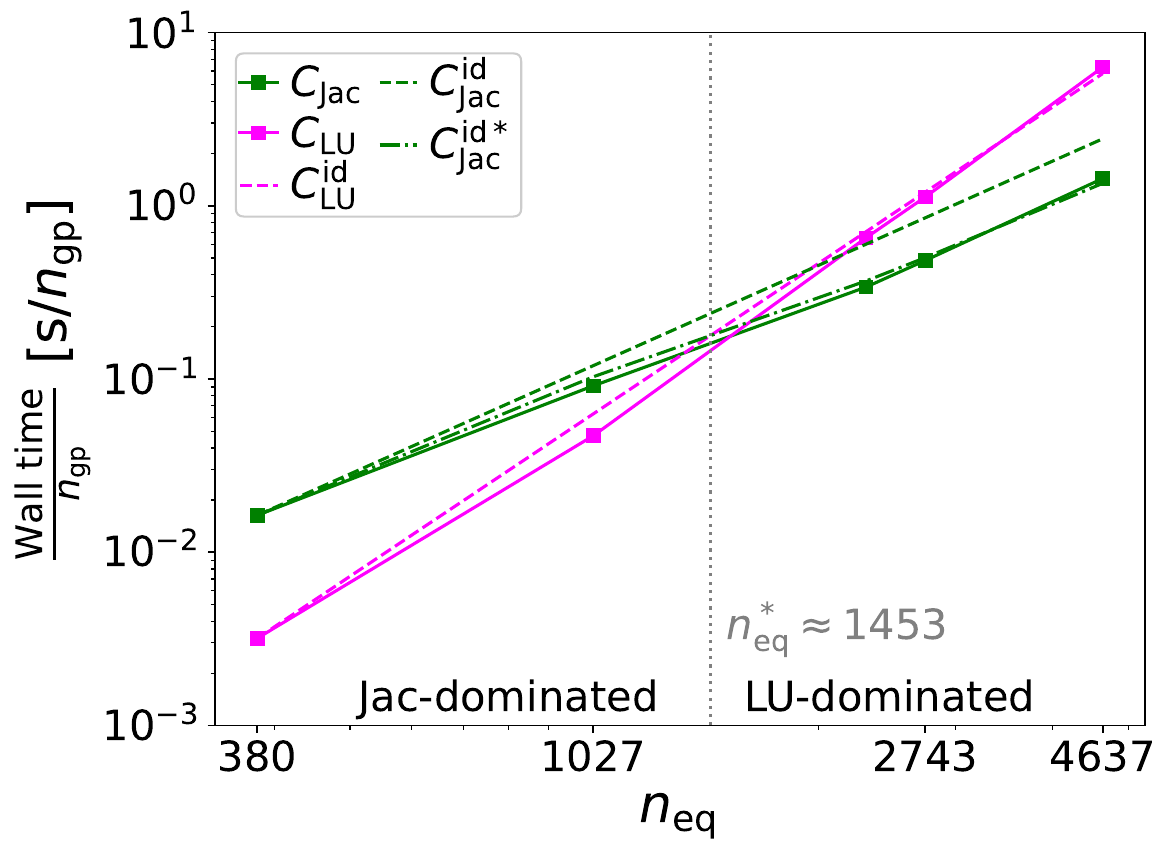}
\caption{Measured wall-clock times per grid point~(solid lines with square markers) for Jacobian evaluation and LU factorization as a function of \(n_{\mathrm{eq}}\).
Dashed lines denote the corresponding theoretical scaling trends. The dash-dotted line represents the Jacobian theoretical scaling trend, which accounts for variations in \(n_{\mathrm{R}}/n_{\mathrm{s}}\) across the kinetic models.
The vertical dotted line marks the predicted crossover at \(n_{\mathrm{eq}}^{*} \approx 1453\).}
\label{fig:idt_adj_scaling}
\end{figure}

Figure~\ref{fig:idt_adj_scaling} reports the cost of Jacobian evaluation and LU factorization, normalized by the number of grid points, together with the corresponding theoretical scaling trends,
\begin{align}
    C_{\mathrm{Jac}}^{\mathrm{id}} &= K \, n_{\mathrm{eq}}^2,\\
    C_{\mathrm{LU}}^{\mathrm{id}}  &= K_2 \, n_{\mathrm{eq}}^3.
\end{align}
The prefactors \(K\) and \(K_2\) are calibrated at the smallest model size by normalizing the measured cost per grid of the smallest benchmark case to the corresponding theoretical scaling law. The obtained prefactors account for both floating-point operations and implementation-dependent overhead.

The measured LU factorization cost closely aligns with the theoretical trend, suggesting that memory-related overheads do not significantly affect the observed scaling behavior. In contrast, the agreement is less pronounced for the Jacobian evaluation, indicating that implementation- or model-dependent effects, which constant prefactors cannot capture, contribute to the deviation of the measured computational cost from the theoretical trend.

A better agreement is observed when the scaling law explicitly includes the variation in the average number of reactions per species, \(n_{\mathrm{R}}/n_{\mathrm{s}}\), rather than absorbing it into the global constant~\(K\). This observation suggests that variations in reaction network connectivity explain the discrepancy between the measured Jacobian evaluation cost and the simpler scaling law.

The observed trend follows from the computation of the numerical Jacobian. Numerical differentiation requires \(n_{\mathrm{eq}}\) evaluations of \(\boldsymbol{F}^{\mathrm{hr}}\), each of which is dominated by reaction rate computation and species production-rate evaluation. Both operations scale approximately linearly with the number of reactions, which yields the more accurate theoretical scaling trend
\begin{equation}
C_{\mathrm{Jac}}^{\mathrm{id*}} = K_1 \, n_{\mathrm{R}}\, n_{\mathrm{eq}}.
\end{equation}
Here, \(K_1\) is a nearly constant prefactor that accounts for the evaluation of reaction rate coefficients and the species production rates. 
Its value depends on the distribution of reaction types in the kinetic model and on the average number of reactants and products per reaction, both of which vary only weakly across the benchmark models.
Considering that \(n_{\mathrm{R}} \sim (n_{\mathrm{R}}/n_{\mathrm{s}})\,n_{\mathrm{s}}\), and that the number of species in the kinetic model \(n_{\mathrm{s}}\) is approximately equal to \(n_{\mathrm{eq}}\), the theoretical Jacobian cost per grid point can be expressed as
\begin{equation}
    \label{eq:scaling_Jacobian}
    C_{\mathrm{Jac}}^{\mathrm{id}*}
    =
    K_1\,\frac{n_{\mathrm{R}}}{n_{\mathrm{s}}}\,n_{\mathrm{eq}}^2.
\end{equation}

Equating the Jacobian and LU theoretical scaling laws defines a predicted crossover size at which the two contributions become comparable:
\begin{equation}
    \label{eq:crossover}
    n_{\mathrm{eq}}^{*} \sim \frac{K_1}{K_2}\,
    \frac{n_{\mathrm{R}}}{n_{\mathrm{s}}}
    \approx 1453.
\end{equation}
Using the prefactors inferred from Fig.~\ref{fig:idt_adj_scaling}, Eq~\eqref{eq:crossover} predicts a crossover between C7-HT-N (\(n_{\mathrm{eq}}=1027\)) and C6 (\(n_{\mathrm{eq}}=2301\)), consistent with the cost decomposition shown in Fig.~\ref{fig:idt_adj_contribution}.
The crossover location is thus determined jointly by the kinetic model, through \(n_{\mathrm{R}}/n_{\mathrm{s}}\), and by implementation-dependent factors through \(K_1\) and \(K_2\). Because smaller models tend to exhibit larger \(n_{\mathrm{R}}/n_{\mathrm{s}}\), the crossover to LU-dominated scaling occurs at larger values of \(n_{\mathrm{eq}}\), extending the regime in which Jacobian evaluation dominates.
Improving adjoint performance, therefore, requires lowering the ratio \(K_1/K_2\), most directly by reducing \(K_1\).

Characterizing the scaling trends of the computational costs of both the
adjoint and forward sensitivity analyses quantifies the speedup trend of the current implementation and enables comparison with the theoretical
estimate. To this end, the total adjoint cost scaling is obtained by
combining the expressions for the two dominant contributions as
\begin{equation}
    \label{eq:scaling_adj_total}
    C_{\mathrm{adj}}^{\mathrm{id}}
    \sim
    C_{\mathrm{Jac}}^{\mathrm{id}} + C_{\mathrm{LU}}^{\mathrm{id}}
    \sim
    K_2\, n_{\mathrm{eq}}^3
    \left(
    \frac{K_1}{K_2}\,
    \frac{n_{\mathrm{R}}}{n_{\mathrm{s}}\,n_{\mathrm{eq}}} + 1
    \right).
\end{equation}

\subsubsection{Forward performance and adjoint speedup analysis}

\begin{figure}[!h]
\centering
\includegraphics[width=\linewidth]{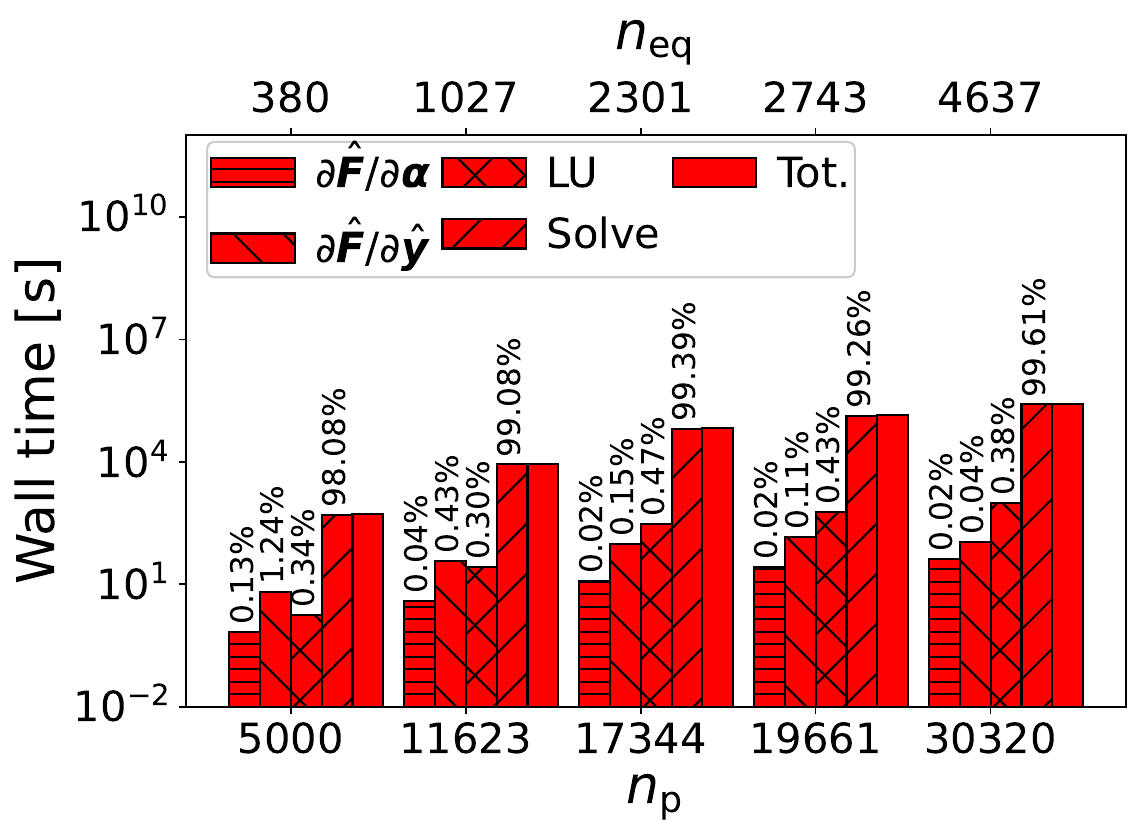}
\caption{Relative contribution of the main computational operations to the total forward computational cost for each benchmarked kinetic model. The label ``Solve'' denotes the cost associated with multiple forward-backward substitutions.
}
\label{fig:idt_forw_contribution}
\end{figure}

Figure~\ref{fig:idt_forw_contribution} shows the relative contribution of the dominant operations to the total wall-clock time for each considered kinetic model with the forward approach. In contrast to the adjoint formulation, the cost of the forward approach is dominated by the repeated forward-backward substitutions, which account for more than~98\% of the total forward time for all considered kinetic models.
This trend is observed even though the number of parameters considered here constitutes only a small portion of the parameters that could be included in the sensitivity analysis.

Therefore, the overall theoretical scaling of the forward cost, normalized by the number of grid points, is
\begin{equation}
    \label{eq:scaling_forw}
    C_{\mathrm{forw}}^{\mathrm{id}} = K_3\, n_{\mathrm p}\, n_{\mathrm{eq}}^2.
\end{equation}
Here, \(K_3\) is again a prefactor calibrated at the smallest model size, accounting for both arithmetic and memory-access operations in the solve routine.

\begin{figure}[!h]
\centering
\includegraphics[width=\linewidth]{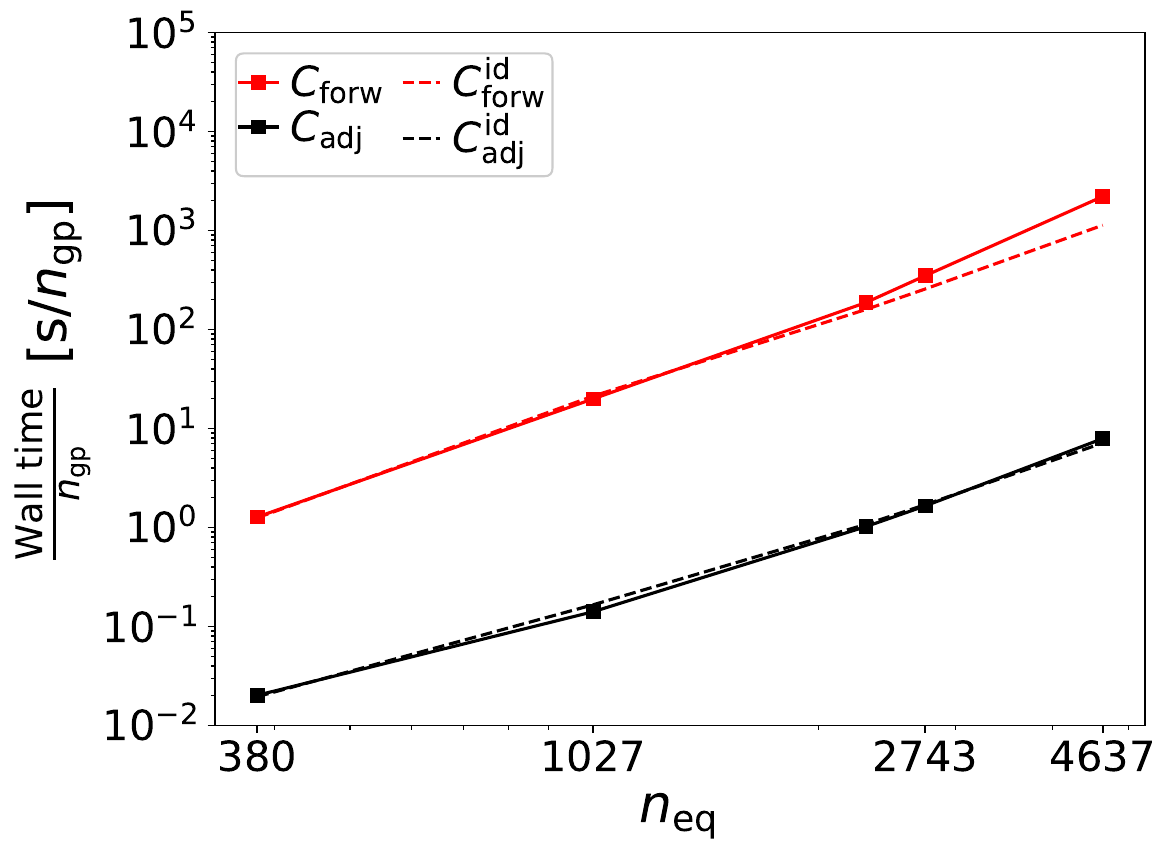}
\caption{Measured wall-clock time per grid point for the adjoint (black solid line) and forward (red solid line) methods as a function of \(n_{\mathrm{eq}}\). Dashed lines denote the corresponding theoretical scaling trends.
All five benchmark cases are included; the tick label at \(n_{\mathrm{eq}}=2301\) (C6) is omitted to avoid overlap with the adjacent label at \(n_{\mathrm{eq}}=2743\).}
\label{fig:idt_scaling_tot}
\end{figure}

Figure~\ref{fig:idt_scaling_tot} compares the measured adjoint and forward wall-clock times, normalized by the number of grid points, with the corresponding theoretical trends. Measurement times for both methods follow the predicted scaling reasonably well over most of the benchmark cases.
From C7 (\(n_{\mathrm{eq}}=2743\)) onward, however, the normalized forward cost begins to diverge from the theoretical trend in Eq.~\eqref{eq:scaling_forw}, whereas the adjoint cost continues to follow the predicted scaling law.
This deviation may be associated with hardware effects at the largest problem sizes, such as reduced cache locality or cross-socket memory traffic during the repeated triangular solve calls.

\begin{figure}[!h]
\centering
\includegraphics[width=\linewidth]{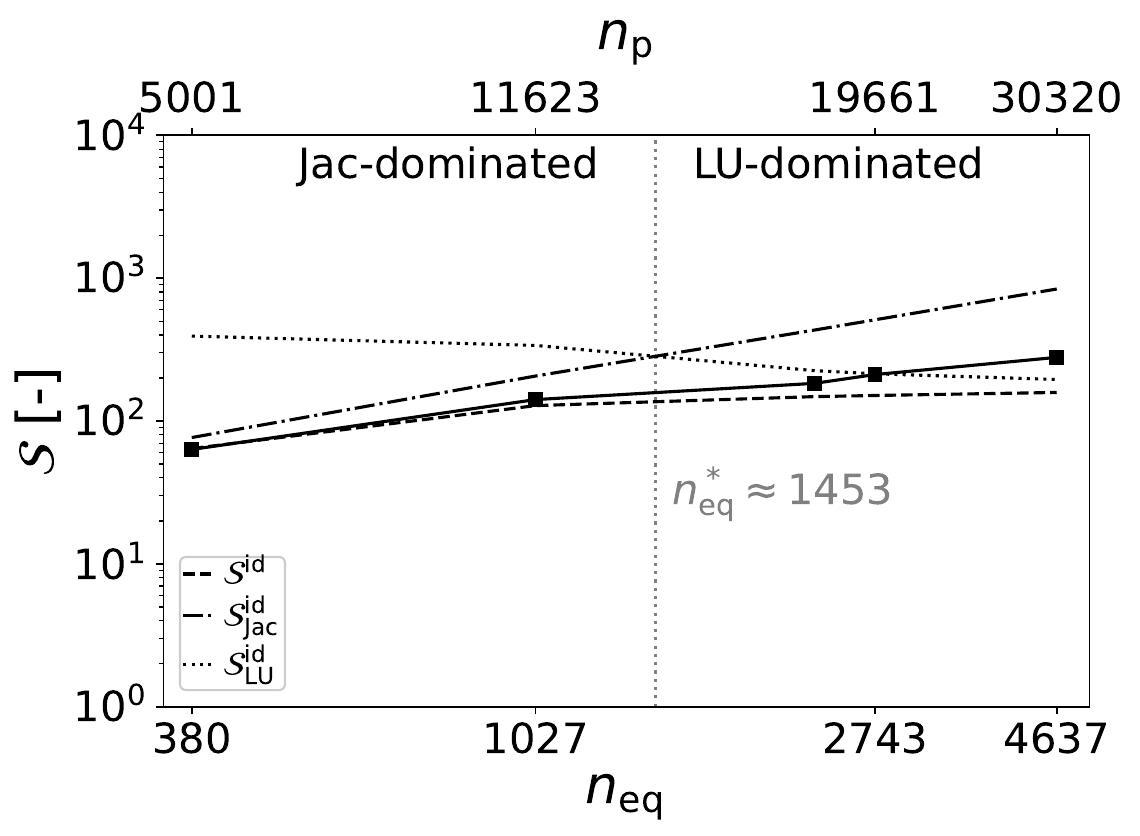}
\caption{Measured speedup (solid line) compared against the expected speedup (dashed), the Jacobian-dominated asymptote (dash-dotted), and the LU-dominated asymptote (dotted).
\(n_{\mathrm{eq}}^{*} \approx 1453\), obtained from Eq.~\eqref{eq:crossover}, marks the transition between the Jacobian-dominated and LU-dominated regimes.
All five benchmarked cases are included; the tick label at \(n_{\mathrm{eq}}=2301\) (C6) is omitted to avoid overlap with the adjacent label at \(n_{\mathrm{eq}}=2743\).}
\label{fig:idt_scaling_speedup}
\end{figure}

Combining Eqs.~\eqref{eq:scaling_adj_total} and~\eqref{eq:scaling_forw}, the theoretical speedup can be expressed as
\begin{equation}
\label{eq:scaling_speedup}
\mathcal{S}^{\mathrm{id}}
=
\frac{C_{\mathrm{forw}}^{\mathrm{id}}}{C_{\mathrm{adj}}^{\mathrm{id}}}
=
\frac{K_3\, n_{\mathrm p}\, n_{\mathrm{eq}}^2}
{K_2\, n_{\mathrm{eq}}^3
\left(\dfrac{K_1 n_{\mathrm{R}}}{K_2 n_{\mathrm{s}}\, n_{\mathrm{eq}}} + 1\right)}.
\end{equation}
This expression admits two limiting asymptotes, corresponding to the Jacobian- and LU-dominated regimes:
\begin{equation}
\mathcal{S}_{\mathrm{Jac}}^{\mathrm{id}} \longrightarrow \frac{K_3}{K_1}\,\frac{n_{\mathrm p}, n_{\mathrm{s}}}{n_{\mathrm{R}}}
\quad\text{and}\quad
\mathcal{S}_{\mathrm{LU}}^{\mathrm{id}} \longrightarrow \frac{K_3}{K_2}\,\frac{n_{\mathrm p}}{n_{\mathrm{eq}}}.
\end{equation}
Figure~\ref{fig:idt_scaling_speedup} compares the measured speedup reported in Tab.~\ref{tab:cpu-comparison} with the trends predicted by the above expression. The Jacobian- and LU-dominated asymptotes are shown by the dashed-dotted and dotted lines, respectively.

Notably, the measured speedup follows the total speedup trend of Eq.~\eqref{eq:scaling_speedup} over most of the benchmark cases: it aligns with the Jacobian-dominated branch where Jacobian evaluation dominates the adjoint cost, then transitions toward a plateau near the LU-dominated asymptote once the LU contribution becomes significant, with the crossover at the \(n_{\mathrm{eq}}^{*}\) identified above.
From C7 (\(n_{\mathrm{eq}}=2743\)) onward, however, the measured speedup begins to diverge from the expected trend, reaching 279 for C8 (\(n_{\mathrm{eq}}=4637\)) and exceeding the LU-dominated asymptote. This final upturn is consistent with the faster-than-predicted growth of the forward cost noted in Fig.~\ref{fig:idt_scaling_tot}, which amplifies the adjoint advantage beyond the predicted scaling.

An alternative solution strategy exploiting the block structure of the Jacobian matrix~\cite{golub2013matrix} was also investigated. This strategy reduces the computational cost of both the adjoint and forward sensitivity methods while bringing the adjoint sensitivity-analysis time closer to that of the corresponding primal solution.
Despite these improvements, the speedups between the adjoint and forward methods remain of the same order of magnitude, indicating that the observed gains are not solely attributable to implementation aspects; a detailed description of the block-structured strategy and supporting analysis are provided in the SM.

\section{Conclusions}
This work presents an efficient algorithm for computing the sensitivity coefficients of ignition delay time and extinction strain rate. The proposed approach is based on established reformulations of the governing equations, which enable the quantities of interest to be explicitly included as a component of the system solution vector, thereby allowing the application of direct sensitivity analysis methods.
Based on these reformulations, discrete adjoint equations are derived and implemented for both quantities.

The implementation was verified by comparing adjoint sensitivity coefficients against the corresponding values obtained from the forward and brute-force methods.
Because the proposed method computes formally defined first-order local sensitivities, the dependence on the perturbation factor selection that affects brute-force methods is avoided. 
In addition, combining an arc-length continuation strategy with the adjoint sensitivity analysis method avoids the singularities and numerical difficulties that arise near extinction limits, while enabling efficient sensitivity evaluation even in the vicinity of this critical condition.

A computational complexity analysis establishes the theoretical efficiency gains and provides the reference against which the measured performance is assessed. The performance evaluation shows that adjoint sensitivity analysis achieves one to two orders of magnitude speedup over the forward approach, demonstrating the potential of the adjoint method to substantially accelerate kinetic modeling workflows bottlenecked by sensitivity computation. The benchmarks further localize where that speedup saturates: for the smaller cases, Jacobian evaluation is responsible for the gap to the theoretical ceiling. Reducing its cost constitutes an implementation-specific rather than algorithmic improvement. For larger models, where the LU factorization cost becomes dominant, assessing Jacobian-free Newton--Krylov formulations~\cite{Knoll2004} is a natural next step, trading the per-block cubic cost of a direct factorization for an iterative solve whose per-iteration cost is one order lower in the block size. The benchmark analysis presented here
provides a quantitative reference against which future adjoint
implementations can be evaluated.

\section*{CrediT authorship contribution statement}
\textbf{Francesca Loffredo}: Writing-review \& editing, Writing-original draft, Conceptualization, Software, Validation, Formal analysis. \textbf{Raymond Langer}:  Writing-review \& editing, Writing-original draft, Conceptualization, Software, Validation, Visualization. \textbf{Marco Vivenzo}:  Writing-review \& editing, Conceptualization, Software, Validation, Visualization. \textbf{Sanket Girhe}: Writing-review \& editing, Conceptualization. \textbf{Heinz Pitsch}: Writing–review \& editing, Supervision, Conceptualization, Software.

\section*{Code availability}
The code implementing the proposed methodology can be provided upon request. The authors are actively working to make it publicly available and to facilitate its distribution in the future.

\section*{Declaration of Generative AI and AI-assisted technologies in the
writing process}

During the preparation of this work, the authors used Perplexity.ai and Claude to improve the language and readability of the manuscript, and Cursor with the Claude Sonnet model to assist with coding-related tasks. These tasks included debugging and implementing block-based algorithms for LU factorization and linear solves employed in the direct sensitivity analysis.
All AI-assisted outputs were critically reviewed, verified, and, where necessary, modified by the authors. The authors take full responsibility for the content of this publication.

\section*{Declaration of competing interest}

The authors declare that they have no known competing financial interests or personal relationships that could have appeared to influence the work reported in this paper.

\section*{Acknowledgments}

This work is funded by the European Research Council (ERC) Advanced Grant (ID: 101054894) and the Deutsche Forschungsgemeinschaft (DFG, German Research Foundation) under Germany's Excellence Strategy - Cluster of Excellence 2186/2 "The Integrated Fuel \& Chemical Science Center"- ID: 390919832. The authors further acknowledge the German Federal Ministry of Research, Technology and Space (BMFTR) and the state of North Rhine-Westphalia for supporting this work as part of the NHR funding. Computing resources were granted by RWTH Aachen University, Germany, under project
rwth1433.

\section*{Supplementary material}

Supplementary material associated with this article can be found in a separate PDF file submitted together with the main article.

\FloatBarrier

\bibliographystyle{cnf-num}
\bibliography{cnf-refs}

\end{document}


\newcommand{\supplementarytitle}[4]{%
    \begin{center}
        {\LARGE \textbf{Supplementary Material}}\\[0.5em]
        {\large \textbf{#1}}\\[0.5em]
        #2\\[0.2em]
        {\footnotesize \em #3}\\[-5pt]
        {\footnotesize $^{*}$Corresponding author: \href{mailto:#4}{#4}}
    \end{center}
    \vspace{1em}
}

\supplementarytitle{Efficient sensitivity analysis for combustion problems via reformulation and adjoint methods} %
{Francesca Loffredo$^{*}$, Raymond Langer, Marco Vivenzo, Sanket Girhe, Heinz Pitsch} %
 { Institute for Combustion Technology, RWTH Aachen University, Templergraben 64, 52056 Aachen, Germany} %
{f.loffredo@itv.rwth-aachen.de} %

\section{Introduction}
This supplementary material provides additional details on the methods and analyses behind the main manuscript. It first provides a detailed explanation of the method used to determine the global brute-force perturbation factor. Following, an alternative block-structured solution strategy for the ignition delay time problem is presented, along with performance analyses supporting the main manuscript text. 
Finally, the performance analyses for the ESR problem are also presented.

\section{Selection of brute-force sensitivity analysis perturbation factor}
A global perturbation factor is determined by a stability-region~(``plateau'') detection algorithm applied to each reaction's logarithmic sensitivity coefficient. The plateau region is identified by considering 12 different perturbation factor values,
\begin{equation}
   \gamma -1 =  \varepsilon \in \{1,\,0.8,\,0.7,\,0.5,\,0.2,\,0.1,\,0.05,\,0.01,
                      \,0.005,\,0.001\,,0.0005\,, 0.0001\},
\end{equation}
spanning three orders of magnitude, $10^{-3} \leq \varepsilon \leq 1$. 
Reactions whose mean sensitivity coefficient~($\langle|S_r|\rangle$) falls below the cut-off $\langle|S_r|\rangle < 10^{-6}$ for all the perturbation factors considered are excluded from the analysis.
For the remaining reactions, a pair of consecutive perturbation steps $(\varepsilon_i,\,\varepsilon_{i+1})$ is called ``stable'' for reaction $r$ if the finite-difference slope
\begin{equation}
    t_{i,r} = \frac{S_r(\varepsilon_i) - S_r(\varepsilon_{i+1})}
                   {\log_{10}\varepsilon_i - \log_{10}\varepsilon_{i+1}}
\end{equation}
satisfies $|t_{i,r}| \leq \theta_r$, where the threshold $\theta_r$ was tuned to the 30$th$ percentile of all absolute values of the slopes for that reaction. 

The longest unbroken sequence of stable pairs defines the plateau of reaction $r$. The center of this plateau, $c_r$, is the geometric mean of the pair of midpoints $\tilde\varepsilon_i = \sqrt{\varepsilon_i\,\varepsilon_{i+1}}$,
\begin{equation}
    c_r = \exp\!\left(\frac{1}{L_r}\sum_{i \in \mathcal{P}_r}
          \ln \tilde\varepsilon_i \right),
\end{equation}
where $\mathcal{P}_r$ is the index set covering the plateau and $L_r$ is  its length. We retain only reactions with a plateau length of at least $L_{\min} = 3$.
Among these reactions, not all are equally informative. A reaction with a high sensitivity that remains stable is far more useful for choosing the global perturbation factor than a weakly sensitive reaction with noisy fluctuations. We quantify this with the signal-to-noise ratio (SNR) weight
\begin{equation}
    w_r = \frac{\langle|S_r|\rangle^{2}}
               {\sigma(S_r)}
        \;\simeq\; \langle|S_r|\rangle \cdot \frac{1}{\mathrm{CV}(S_r)},
    \qquad
    \mathrm{CV}(S_r) = \frac{\sigma(S_r)}{\langle|S_r|\rangle},
\end{equation}
where $\langle|S_r|\rangle$ is the mean of $|S_r(\varepsilon)|$ over all runs, $\sigma(S_r)$ is its standard deviation, and \(\mathrm{CV}\) is the coefficient of variation, which measures the stability of sensitivity coefficients.
Let $\{c_r\}_{r=1}^{N}$ be the plateau centers of the $N$ qualifying reactions and $\bar{w}_r = w_r / \sum_r w_r$ the normalized SNR weights. The recommended perturbation factor is the SNR-weighted geometric mean
\begin{equation}
    \varepsilon^* = \exp\!\left(\sum_{r=1}^{N} \bar{w}_r \ln c_r\right).
\end{equation}

A key limitation is that the approach relies on user-defined threshold parameters, which need tuning to obtain physically meaningful results.

\section{Block-tridiagonal solver for the sensitivity Jacobian}

The structure of the Jacobian~(\(A=\partial\hat{\boldsymbol{F}}^{\mathrm{hr}}/\partial\hat{\boldsymbol{y}}^{\mathrm{hr}}\)) is block-tridiagonal and is given as,
\begin{equation}
A=\frac{\partial\hat{\boldsymbol{F}}^{\mathrm{hr}}}{\partial\hat{\boldsymbol{y}}^{\mathrm{hr}}}=\begin{pmatrix}
D_0 & U_0 & & \\
L_1 & D_1 & U_1 & \\
 & \ddots & \ddots & \ddots \\
 & & L_{n_{\mathrm{gp}}-1} & D_{n_{\mathrm{gp}}-1}
\end{pmatrix},
\end{equation}
where the diagonal blocks \(D_{\mathrm{idx}}\in\mathbb{R}^{n_{\mathrm{eq}}\times n_{\mathrm{eq}}}\) are dense, since they gather the local kinetic coupling, while the off-diagonal blocks are sparse. The species and temperature equations use a backward-difference scheme and the \(\tau\)-equation a forward-difference scheme, so that
\begin{equation}
L_{\mathrm{idx}}=-\frac{1}{h_{\mathrm{idx}}}\sum_{k\neq n_\tau}\boldsymbol{e}_k\boldsymbol{e}_k^\top,
\qquad
U_{\mathrm{idx}}=\frac{1}{h_{\mathrm{idx}}}\,\boldsymbol{e}_{n_\tau}\boldsymbol{e}_{n_\tau}^\top ,
\end{equation}
where \(h_{\mathrm{idx}}\) is the local grid spacing and \(\boldsymbol{e}_k\) is the \(k\)-th canonical basis vector of \(\mathbb{R}^{n_{\mathrm{eq}}}\).
Consequently, \(L_{\mathrm{idx}}\) is diagonal with entries equal to \(-1/h_{\mathrm{idx}}\) in every row except the \(\tau\)-row \(n_\tau\), which is empty, while \(U_{\mathrm{idx}}\) is an outer product with a single nonzero entry, \([U_{\mathrm{idx}}]_{n_\tau n_\tau}=1/h_{\mathrm{idx}}\). %

Both the forward and the adjoint methods require the solution of a block-tridiagonal linear system \(C\boldsymbol{x}=\boldsymbol{b}\) with \(C=A\) and \(C=A^\top\), respectively.

For the forward method, the system is solved through a block LU factorization of the coefficient matrix, \(A=\mathcal{L}\,\mathcal{U}\), i.e., the block Thomas algorithm~\cite{golub2013matrix}.
The block lower triangular factor \(\mathcal{L}\) retains the original sub-diagonal blocks \(L_{\mathrm{idx}}\) and carries the updated blocks \(D_{\mathrm{idx}}'\) on its main diagonal, while the block upper triangular factor \(\mathcal{U}\) has identity blocks on its main diagonal and the scaled blocks \((D_{\mathrm{idx}}')^{-1}U_{\mathrm{idx}}\) on its upper diagonal,
\begin{equation}
\mathcal{L}=\begin{pmatrix}
D_0' & & & \\
L_1 & D_1' & & \\
 & \ddots & \ddots & \\
 & & L_{n_{\mathrm{gp}}-1} & D_{n_{\mathrm{gp}}-1}'
\end{pmatrix},
\qquad
\mathcal{U}=\begin{pmatrix}
I & (D_0')^{-1}U_0 & & \\
 & I & (D_1')^{-1}U_1 & \\
 & & \ddots & \ddots \\
 & & & I
\end{pmatrix},
\end{equation}
where each \(D_{\mathrm{idx}}'\) is obtained from the forward elimination as:
\begin{equation}
D_0'=D_0,\qquad
D_{\mathrm{idx}}'=D_{\mathrm{idx}}-L_{\mathrm{idx}}\,(D_{\mathrm{idx}-1}')^{-1}\,U_{\mathrm{idx}-1}\quad(\mathrm{idx}\ge 1).
\end{equation}
Since \(U_{\mathrm{idx}-1}=h_{\mathrm{idx}-1}^{-1}\boldsymbol{e}_{n_\tau}\boldsymbol{e}_{n_\tau}^\top\) has a single nonzero entry, the triple product collapses to
\begin{equation}
L_{\mathrm{idx}}\,(D_{\mathrm{idx}-1}')^{-1}\,U_{\mathrm{idx}-1}
=\frac{1}{h_{\mathrm{idx}-1}}\,\big(L_{\mathrm{idx}}\,\boldsymbol{w}_{\mathrm{idx}-1}\big)\,\boldsymbol{e}_{n_\tau}^\top ,
\qquad
\boldsymbol{w}_{\mathrm{idx}-1}:=(D_{\mathrm{idx}-1}')^{-1}\boldsymbol{e}_{n_\tau},
\end{equation}
so that the update of a diagonal block requires one triangular solve for the vector \(\boldsymbol{w}_{\mathrm{idx}-1}\). 
For the adjoint method, the same argument applies to the transposed Jacobian with the outer product acting on the left, in a way that the correction affects a single row.

The final solution \(\boldsymbol{x}\) is obtained in two solve sweeps, \(\mathcal{L}\boldsymbol{z}=\boldsymbol{b}\) followed by \(\mathcal{U}\boldsymbol{x}=\boldsymbol{z}\).
The first system is solved forward, from \(\mathrm{idx}=0\) to \(\mathrm{idx}=n_{\mathrm{gp}}-1\),
\begin{equation}
\boldsymbol{z}_0=(D_0')^{-1}\boldsymbol{b}_0, \qquad
\boldsymbol{z}_{\mathrm{idx}}=(D_{\mathrm{idx}}')^{-1}\!\big(\boldsymbol{b}_{\mathrm{idx}}-L_{\mathrm{idx}}\,\boldsymbol{z}_{\mathrm{idx}-1}\big)\quad(\mathrm{idx}\ge 1),
\end{equation}
where each small dense block \(D_{\mathrm{idx}}'\) is factorized once with a dense LU decomposition, and its factors are reused both for the auxiliary vector \(\boldsymbol{w}_{\mathrm{idx}}=(D_{\mathrm{idx}}')^{-1}\boldsymbol{e}_{n_\tau}\) and for the right-hand side solve. The second system is then solved backward, from \(\mathrm{idx}=n_{\mathrm{gp}}-1\) down to \(\mathrm{idx}=0\),
\begin{equation}
\boldsymbol{x}_{n_{\mathrm{gp}}-1}=\boldsymbol{z}_{n_{\mathrm{gp}}-1},\qquad
\boldsymbol{x}_{\mathrm{idx}}=\boldsymbol{z}_{\mathrm{idx}}-(D_{\mathrm{idx}}')^{-1}U_{\mathrm{idx}}\,\boldsymbol{x}_{\mathrm{idx}+1}
=\boldsymbol{z}_{\mathrm{idx}}-\frac{x_{\tau,\,\mathrm{idx}+1}}{h_{\mathrm{idx}}}\,\boldsymbol{w}_{\mathrm{idx}} ,
\end{equation}
where \(x_{\tau,\,\mathrm{idx}+1}\) is the \(\tau\)-component of the next block. 

\subsection{Block-structured solution strategy: cost breakdown and scaling}
\label{sec:sm_block}

In this section, the computational-cost analysis is repeated for both the adjoint and the forward method using the variant whose linear-algebra operations exploit the block structure of the Jacobian matrix~\cite{golub2013matrix}. This strategy carries out both the LU factorization and the forward–backward substitutions on smaller dense blocks, which particularly benefits the forward method: since it requires one linear solve per parameter, the reduced per-solve cost is amplified across all solves. As before, the benchmarks were run in single-threaded mode on the CLAIX-2023-HPC system at RWTH Aachen University, equipped with two Intel Xeon Platinum~8468 Sapphire Rapids processors (96 cores in total) and 256~GB of main memory. The code was compiled with the Intel oneAPI toolchain using performance-oriented flags (\texttt{-O3}, \texttt{-ffast-math}, \texttt{-DNDEBUG}). In the block strategy, the LU factorization and the forward–backward substitutions are performed with Eigen's~\cite{eigen} built-in dense LU decomposition, applied to the individual dense blocks.

Table~\ref{tab:sm_costs} collects the total wall-clock time of the forward and adjoint sensitivity analyses for both the reference and the block-structured implementations. Measurements are again collected by determining sensitivity coefficients for a stoichiometric~\ce{CH4}/\ce{air} mixture at \SI{1250}{K} and atmospheric pressure, simulated using kinetic models of increasing size extracted from C3MechV4.0.1~\cite{c31,c32}. The wall-clock time for solving the primal problem is reported alongside for each kinetic model as a reference.
The block strategy consistently reduces the cost of both methods by up to $\approx 2.7\times$ for the adjoint and $\approx 3.6\times$ for the forward at the largest problem size. More importantly, the block adjoint-to-primal cost ratio is lower than the reference implementation counterpart and remains between roughly 1.7 and 3.8, confirming that the block strategy brings the adjoint sensitivity cost close to that of the primal solution.

\begin{table}[htbp]
  \centering
  \caption{Wall-clock times for computing \(S_{\tau,\alpha_i}\) for \(n_{\mathrm p}\) parameters with the forward and adjoint methods, comparing the reference and block-structured~(block) implementations; the cost of the primal solution is also reported.
Conditions: \ce{CH4}/\ce{air}, \(\phi = 1\), \SI{1250}{K}, \SI{1}{atm}. Labels with HT denote kinetic models that include only high-temperature chemistry, while N indicates the inclusion of the \ce{N}- and N-C-chemistry submodule. The lower panel reports the reference-to-block reduction factor for each method and both the adjoint-to-primal and block adjoint-to-primal cost ratios.}
  \label{tab:sm_costs}
  \footnotesize
  \setlength{\tabcolsep}{10pt}
  \begin{tabular}{lccccc}
    \hline
    \textbf{Mech.}             & \textbf{C4-HT} & \textbf{C7-HT-N} & \textbf{C6} & \textbf{C7} & \textbf{C8} \\
    \textbf{$n_{\mathrm p}$}   & 5001            & 11623              & 17344       & 19661       & 30320       \\
    \textbf{$n_{\mathrm{eq}}$} & 380             & 1027               & 2301        & 2743        & 4637        \\
    \hline
    \multicolumn{6}{c}{\textbf{Wall time [s]}}\\
    \hline
    Time primal               & 2.50 & 15.0 & 58.8  & 83.1   & 205   \\
    Time adj.                 & 8.24 & 61.7  & 358   & 643    & 925    \\
    Time adj. block           & 7.52 & 48.1  & 193   & 315    & 345    \\
    Time forw.                & 521  & 8720  & 65700 & 136000 & 258000 \\
    Time forw. block          & 183  & 3610  & 19600 & 38700  & 72400  \\
    \hline
     \multicolumn{6}{c}{\textbf{Cost ratio [-]}}
     \\
    \hline
    (Time adj.) / (Time adj. block)      & 1.10 & 1.28 & 1.85 & 2.04 & 2.68 \\
    (Time forw.) / (Time forw. block)    & 2.85 & 2.42 & 3.35 & 3.51 & 3.56 \\
    (Time adj.) / (Time primal)          & 3.30 & 4.11 & 6.09 & 7.73 & 4.51 \\
    (Time adj. block) / (Time primal)    & 3.01 & 3.21 & 3.28 & 3.79 & 1.68 \\
    (Time forw. block) / (Time adj. block)  & 24.3 & 75.1 & 102 & 123 & 210 \\
    \hline
  \end{tabular}
\end{table}

\begin{figure}[htbp]
  \centering
  \begin{subfigure}[t]{0.49\linewidth}
    \centering
    \includegraphics[width=\linewidth]{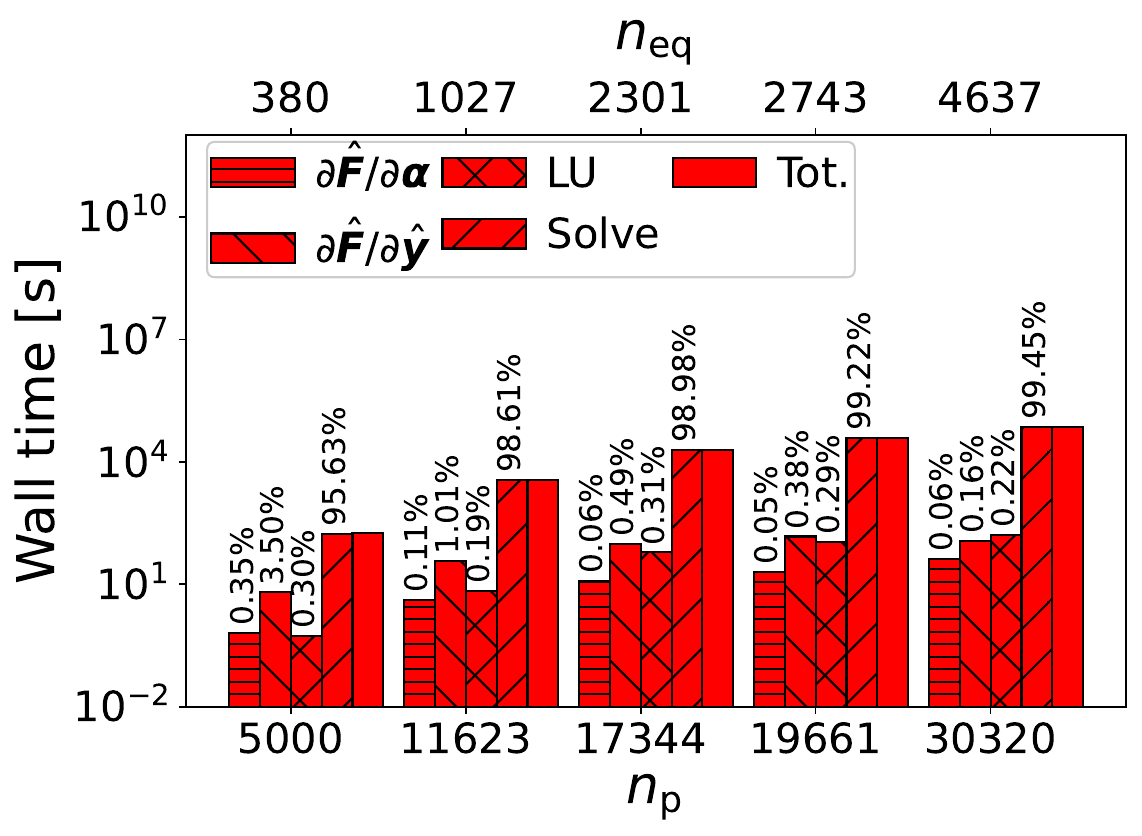}
    \caption{}
  \end{subfigure}
  \hfill
  \begin{subfigure}[t]{0.49\linewidth}
    \centering
    \includegraphics[width=\linewidth]{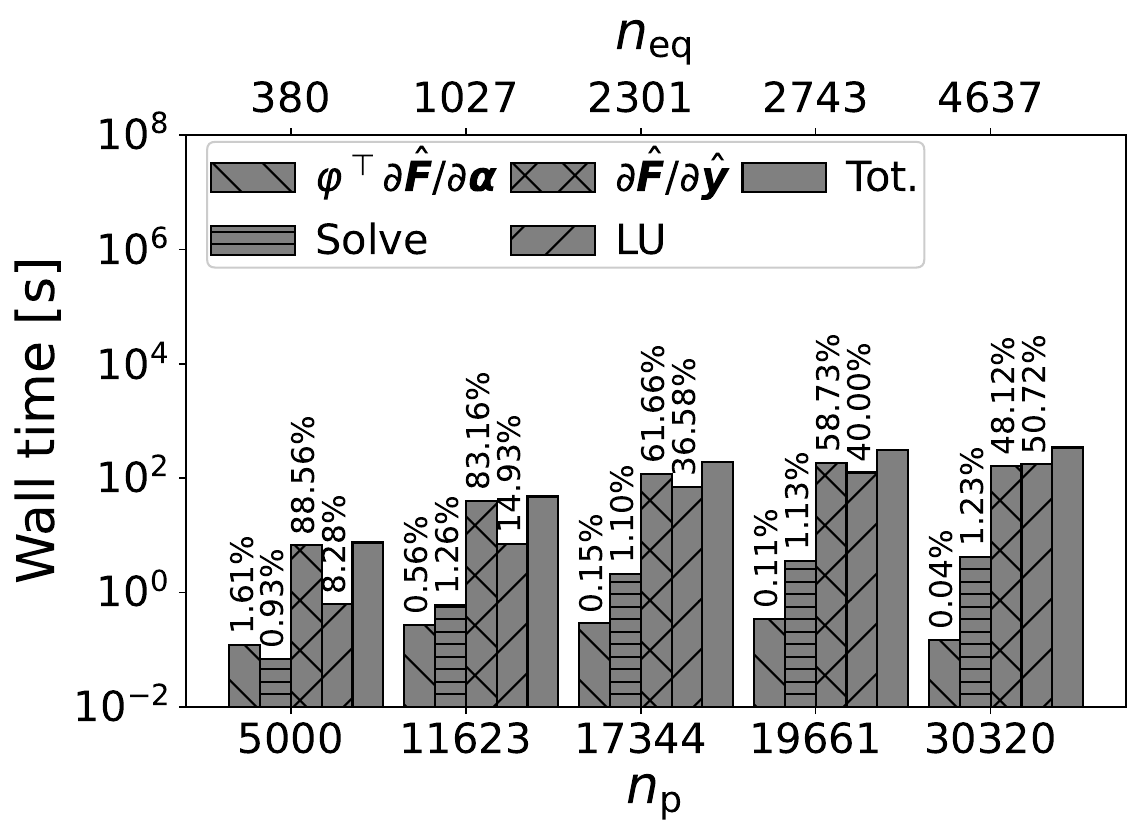}
    \caption{}
  \end{subfigure}
  \caption{(a) Relative contribution of the main computational operations to the total forward (block-structured strategy) cost for each benchmarked kinetic model.~(b) Relative contribution of the main computational operations to the total adjoint (block-structured strategy) cost for each benchmarked kinetic model.}
 \label{fig:forw_adj_breakdown_block}
\end{figure}

The effect of the more efficient linear-algebra strategy on the operation-level cost breakdown is examined first.
Figure~\ref{fig:forw_adj_breakdown_block} reports the per-component wall-time breakdown for both the forward and adjoint sensitivity analyses as a function of problem size. 
For the forward sensitivity analysis, the repeated forward--backward substitution remains dominant across all problem sizes, accounting for $\approx 96\%$ to $\approx 100\%$ of the cost. For the adjoint method, as in the reference case, the total cost is instead dominated by the Jacobian evaluation and the LU factorization, but the improved strategy pushes their crossover toward the largest model, delaying entry into the LU-limited regime.

\begin{figure}[htbp]
  \centering
  \includegraphics[width=0.55\linewidth]{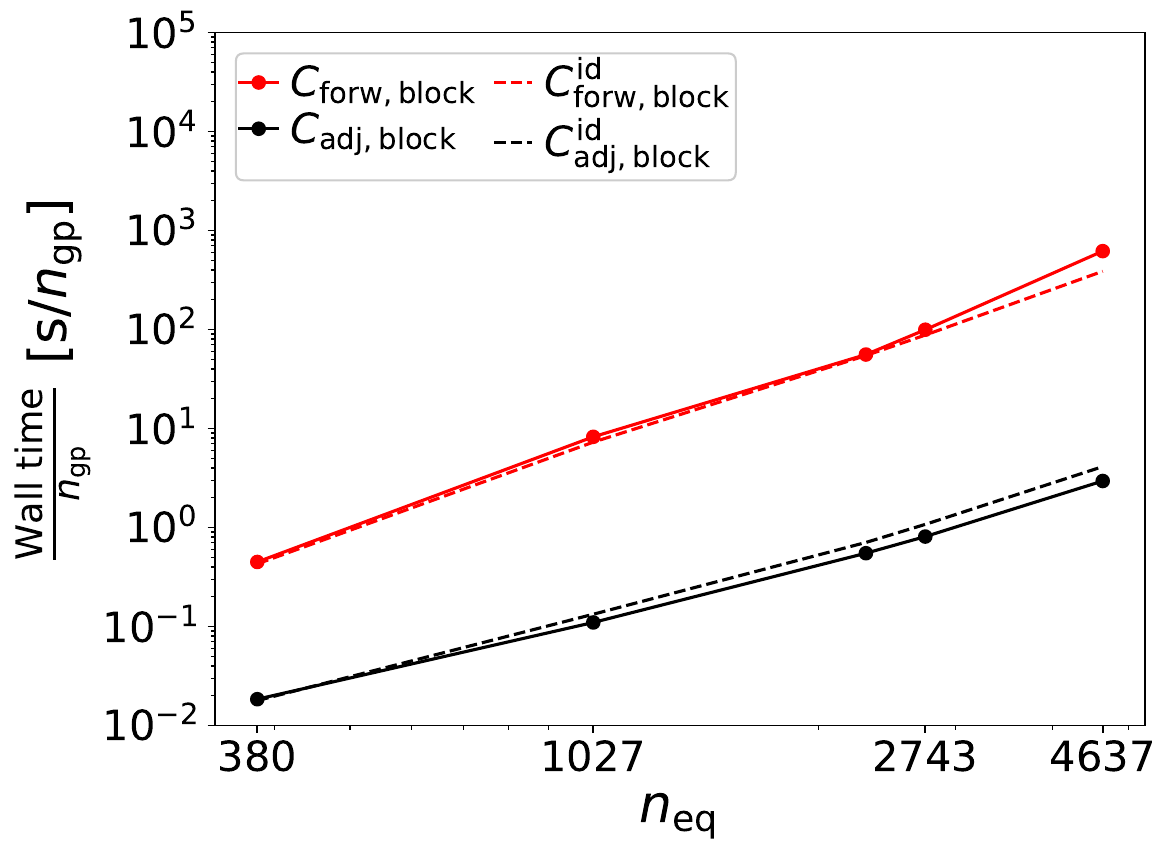}
  \caption{ Measured wall-clock time per grid point for the adjoint (black solid line) and forward (red solid line) methods with the block-structured strategy as a function of \(n_{\mathrm{eq}}\). Dashed lines denote the corresponding theoretical scaling trends; the tick label at \(n_{\mathrm{eq}}=2301\) (C6) is omitted to avoid overlap with the adjacent label.}
  \label{fig:sm_scaling}
\end{figure}

Figure~\ref{fig:sm_scaling} compares the measured scaling of the forward and adjoint costs using the block-structured strategy, normalized per grid point, with the corresponding ideal theoretical estimates ($C_{\mathrm{forw,block}}^{\mathrm{id}}$ and $C_{\mathrm{adj,block}}^{\mathrm{id}}$).
The theoretical scaling laws follow the same equations reported in the main text, but with a different prefactor, calibrated to the block-structure strategy measured at the smallest model size. Both curves track their ideal slopes closely.

\begin{figure}[htbp]
  \centering
  \includegraphics[width=0.55\linewidth]{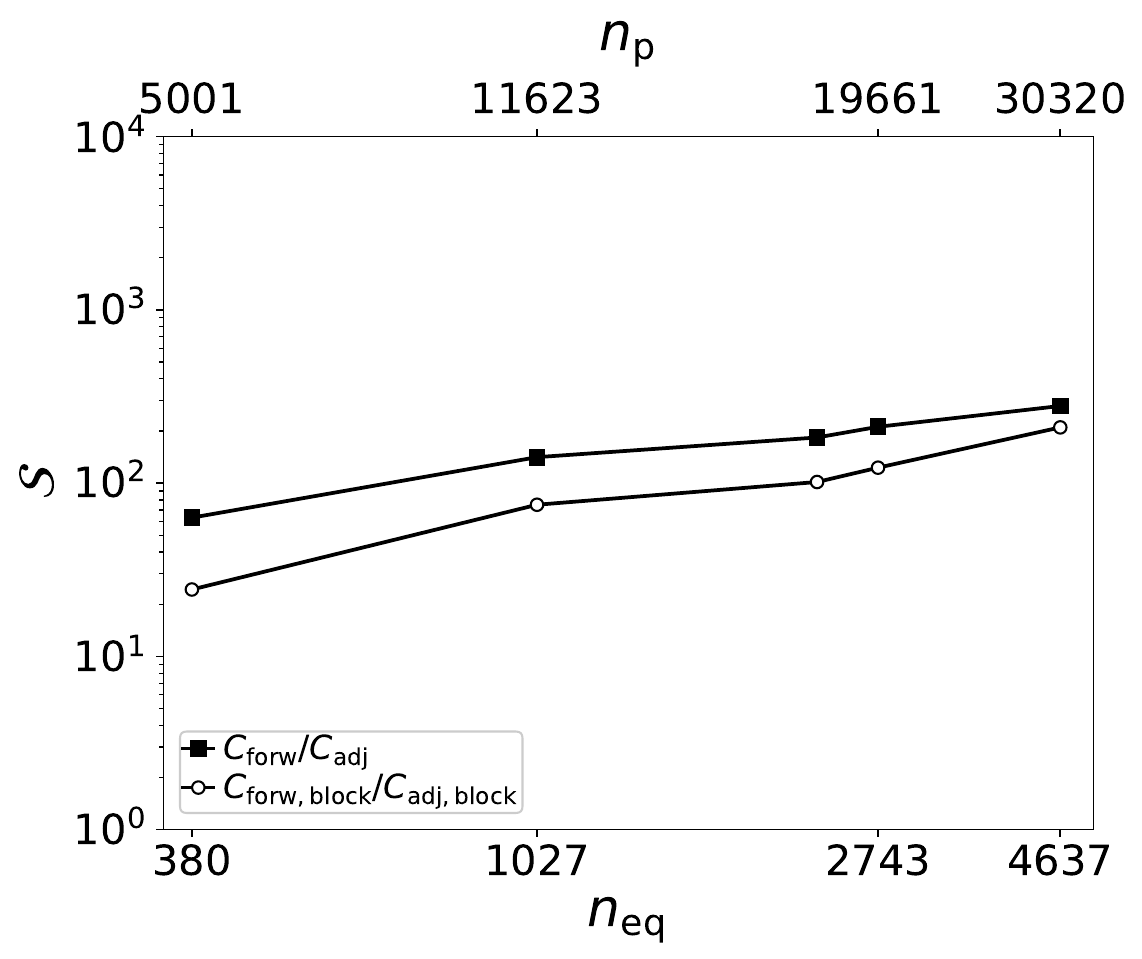}
  \caption{Measured speedup obtained from the reference implementation~(solid line with square) and the block-structured strategy~(solid line with circle); the tick label at \(n_{\mathrm{eq}}=2301\) (C6) is omitted to avoid overlap with the adjacent label.}
  \label{fig:sm_speedup}
\end{figure}

Finally, Fig.~\ref{fig:sm_speedup} reports the speedup for both the reference and the block-structured implementations. Although the block strategy lowers the absolute cost of both methods, the obtained speedup remains within the same order of magnitude, $\mathcal{O}(10^{1})$--$\mathcal{O}(10^{2})$, and follows the same increasing trend.
Therefore, the advantage of the adjoint method can not be attributed solely to implementation aspects.

\newpage

\section{ESR problem performance breakdown}
\begin{figure}[htbp]
  \centering
  \begin{subfigure}[t]{0.49\linewidth}
    \centering
    \includegraphics[width=\linewidth]{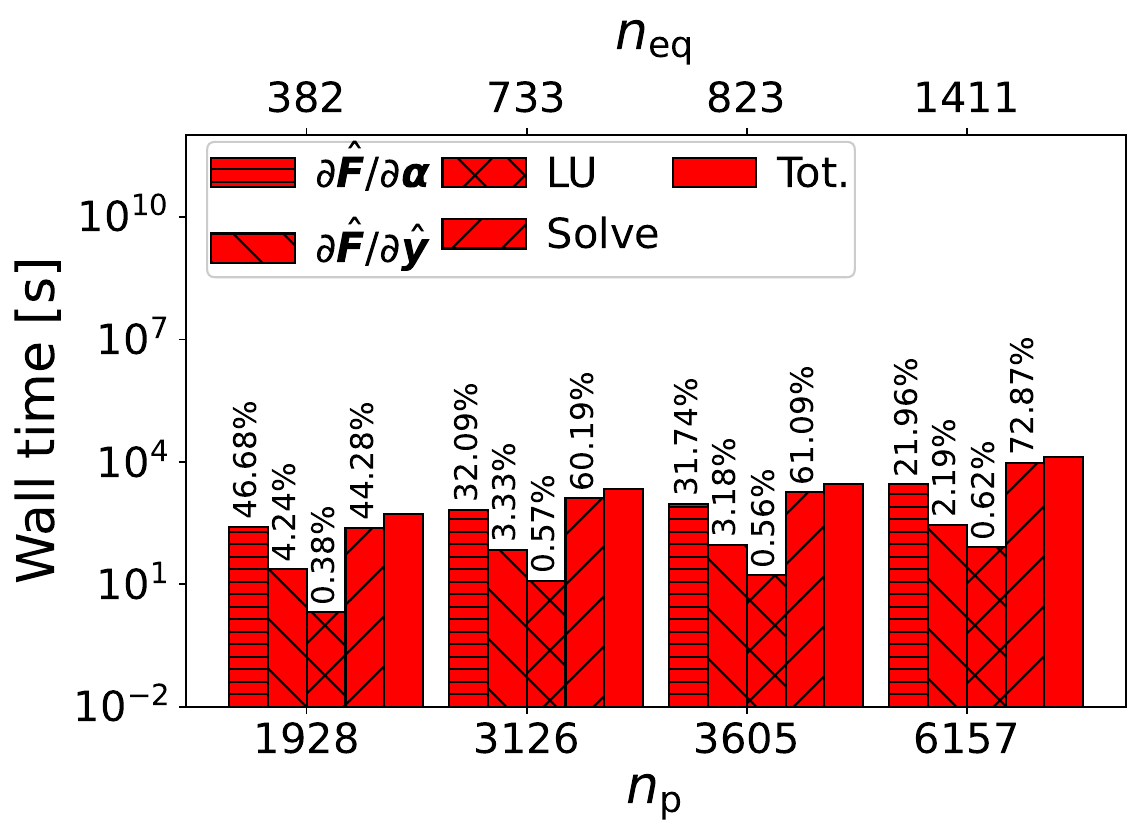}
    \caption{}
  \end{subfigure}
  \hfill
  \begin{subfigure}[t]{0.49\linewidth}
    \centering
    \includegraphics[width=\linewidth]{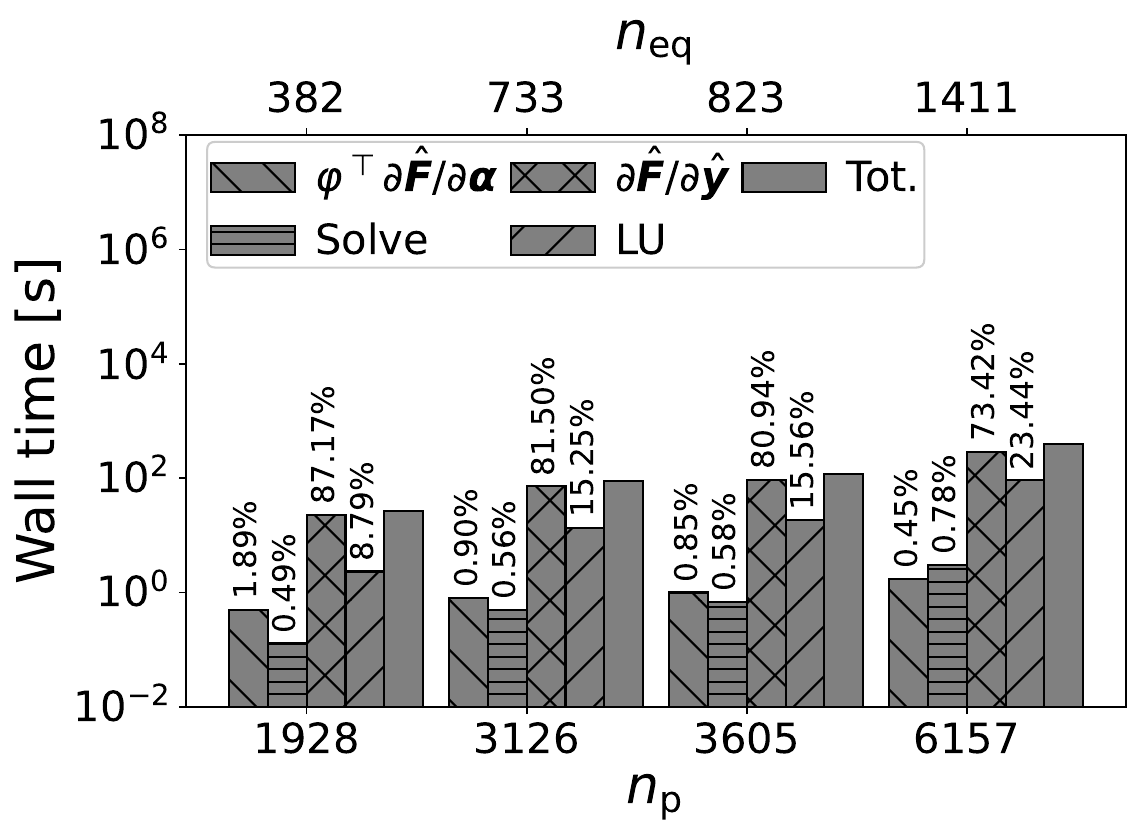}
    \caption{}
  \end{subfigure}
  \caption{(a) Relative contribution of the main computational operations to the total forward cost for each benchmarked kinetic model.~(b) Relative contribution of the main computational operations to the total adjoint cost for each benchmarked kinetic model.}
 \label{fig:forw_adj_breakdown}
\end{figure}

Figure~\ref{fig:forw_adj_breakdown} shows the computational cost breakdown of the forward and adjoint sensitivity analyses for the ESR problem. As in the IDT case, the adjoint method is dominated by the LU factorization and the Jacobian evaluation. Compared to the IDT case, however, the Jacobian computation incurs a higher relative cost, delaying the onset of the LU-dominated regime. This is attributed to the second-order discretization of both the ESR governing equations and the numerical Jacobian, as well as to the presence of transport-related terms in the ESR formulation, all of which increase the computational cost per grid point.

Consistent with the IDT results, in the forward sensitivity analysis, the dominant cost for larger models is the linear solve; however, a substantial portion of the computation is also spent evaluating \(\left[\partial \hat{\boldsymbol{F}} / \partial \boldsymbol{\alpha}\right]\). Here, \(\partial \hat{\boldsymbol{F}} / \partial \alpha_i\) is obtained by algorithmic differentiation, propagating a scalar tangent through the evaluation of \(\hat{\boldsymbol{F}}\).
The kinetic parameters considered in this work enter \(\hat{\boldsymbol{F}}\) only through the reaction rate coefficients and the production rates, whose update is, according to profiling, the dominant cost of the residual evaluation. Since each scalar tangent sweep carries the derivative with respect to a single parameter, this property update is repeated at every grid point once per parameter. Hence, the number of property updates scales proportionally to \(n_{\mathrm{p}}\).
These \(n_{\mathrm{p}}\)-dependent property updates in the forward method stem from using algorithmic differentiation in place of the analytic derivatives, and thus reflect the current implementation rather than an intrinsic limitation of the forward approach.

\begin{figure}[htbp]
  \centering
  \includegraphics[width=0.55\linewidth]{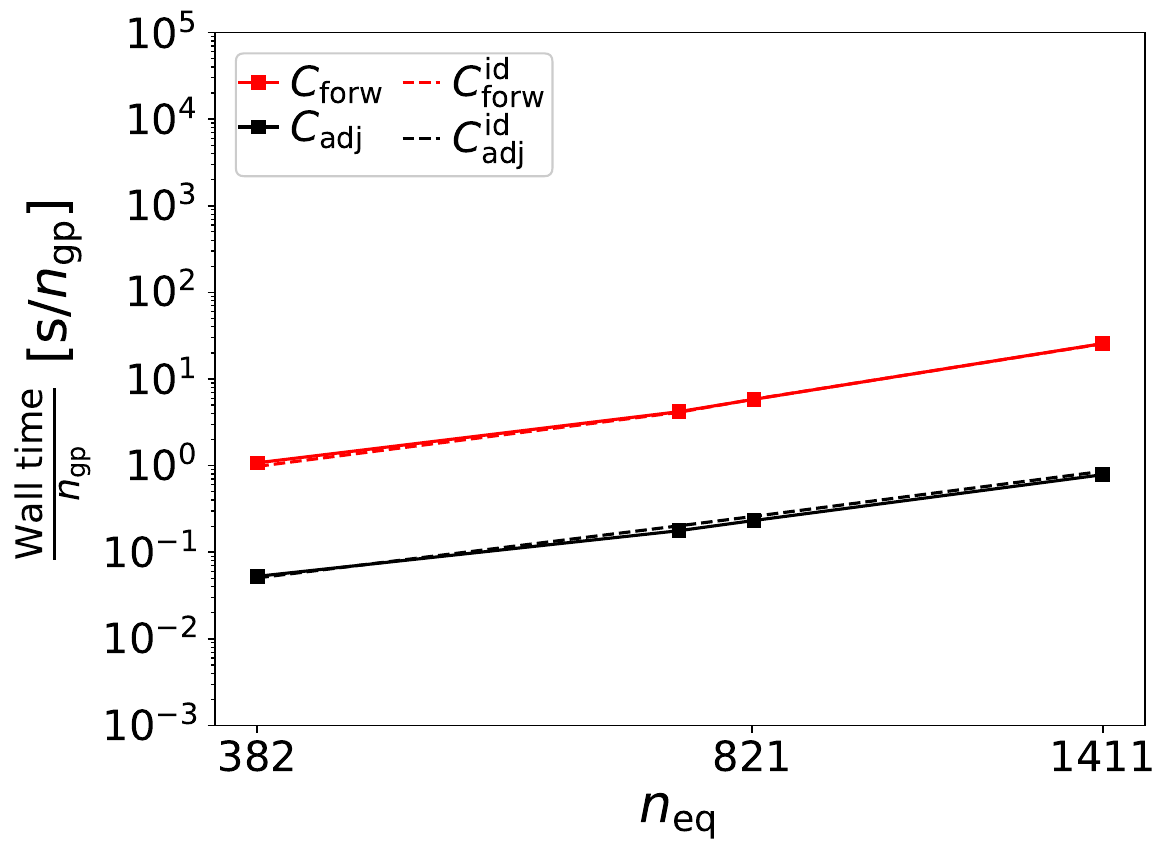}
  \caption{Measured wall-clock time per grid point for the adjoint~(black solid line) and forward~(red solid line) methods as a function of \(n_{\mathrm{eq}}\). Dashed lines denote the corresponding theoretical scaling trends. All four benchmark cases are included; the tick label at \(n_{\mathrm{eq}}=733\)~(C4) is omitted to avoid overlap with the adjacent label.}
  \label{fig:esr_scaling}
\end{figure}

The scaling behaviour of the two methods is quantified by analysing the scaling of their dominant contributions. All costs are again normalized by the number of grid points \(n_{\mathrm{gp}}\), and the constant prefactors are calibrated at the smaller problem size.  
For the adjoint method, the runtime is dominated by the Jacobian evaluation and the LU factorization, which scale as
\begin{align}
    C_{\mathrm{Jac}}^{\mathrm{id}} &= K_4 \, (\frac{n_{\mathrm{R}}}{n_{\mathrm{s}}} +1 )\,n_{\mathrm{eq}}^2,\\
    C_{\mathrm{LU}}^{\mathrm{id}}  &= K_5 \, n_{\mathrm{eq}}^3.
\end{align}
Here, the \(+1\) in \( C_{\mathrm{Jac}}^{\mathrm{id}}\) accounts for the \(\mathcal{O}(n_{\mathrm{eq}}^2)\) contributions per grid point, e.g.\ the recomputation of binary diffusion coefficients upon temperature perturbations.
For the forward method, the cost is instead governed by the \(n_{\mathrm{p}}\) forward--backward solves and the \(n_{\mathrm{p}}\)-proportional updates of the \(\hat{\boldsymbol{F}}\) derivative, which scale as
\begin{align}
    C_{\mathrm{solve}}^{\mathrm{id}} &= K_6 \, n_{\mathrm{p}} \, n_{\mathrm{eq}}^2,\\
    C_{\mathrm{rhs}}^{\mathrm{id}}   &= K_7 \, n_{\mathrm{p}} \, n_{\mathrm{R}}.
\end{align}
Summing the dominant contributions of each method yields the total scaling laws
\begin{align}
    C_{\mathrm{adj}}^{\mathrm{id}}  &= K_4 \, (\frac{n_{\mathrm{R}}}{n_{\mathrm{s}}} +1 )\,n_{\mathrm{eq}}^2 + K_5 \, n_{\mathrm{eq}}^3,\\
    C_{\mathrm{forw}}^{\mathrm{id}} &= K_6 \, n_{\mathrm{p}} \, n_{\mathrm{eq}}^2 + K_7 \, n_{\mathrm{p}} \, n_{\mathrm{R}}.
\end{align}
Figure~\ref{fig:esr_scaling} reports the measured total wall-clock time of the forward and adjoint methods, normalized by \(n_{\mathrm{gp}}\), together with the derived scaling laws; the measured trends follow the predicted scaling trends closely.

\begin{figure}[htbp]
  \centering
  \includegraphics[width=0.55\linewidth]{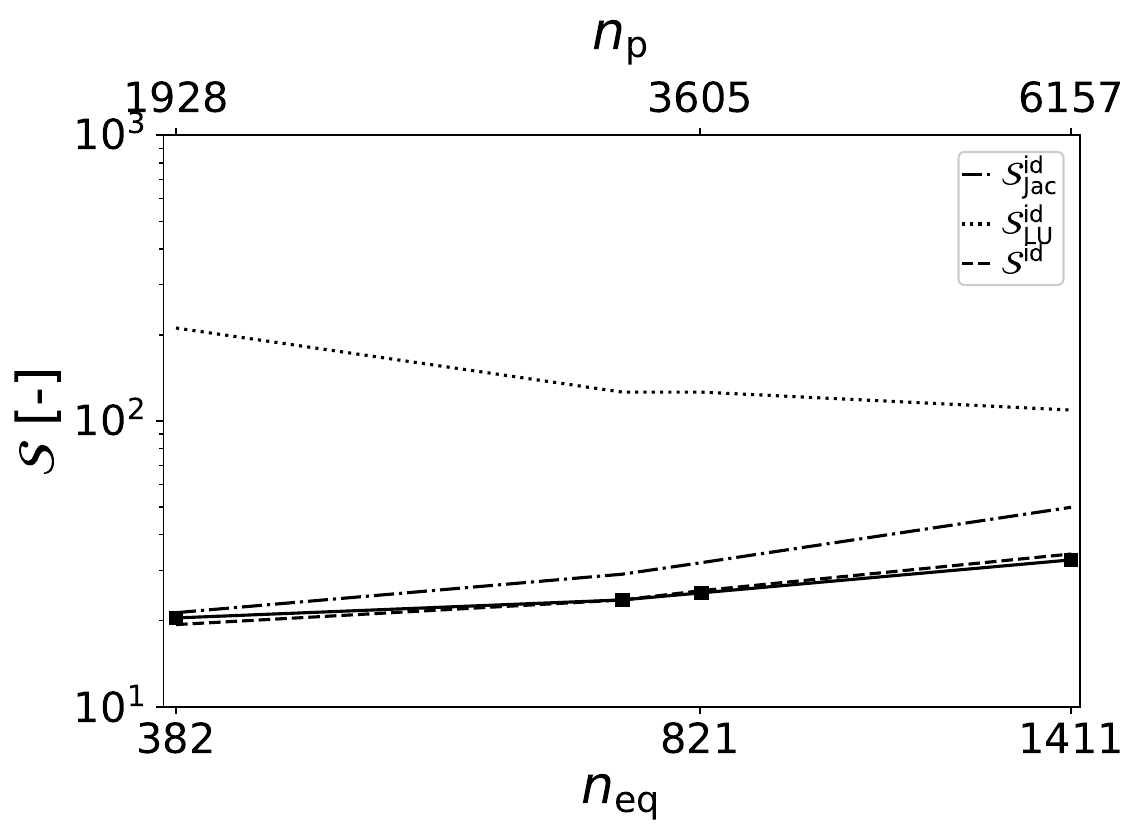}
  \caption{Measured speedup~(solid line) compared against the total speedup scaling law~(dashed), the Jacobian-dominated~(dash-dotted), and the LU-dominated~(dotted) bounds. All four benchmarked cases are included; the tick label at \(n_{\mathrm{eq}}=733\)~(C4) is omitted to avoid overlap with the adjacent label.}
  \label{fig:esr_speedup}
\end{figure}

The obtained speedup scaling law and the two limiting trends are given by
\begin{align}
    \mathcal{S}^{\mathrm{id}}                &= \frac{C_{\mathrm{forw}}^{\mathrm{id}}}{C_{\mathrm{adj}}^{\mathrm{id}}}, \\
    \mathcal{S}_{\mathrm{Jac}}^{\mathrm{id}} &= \frac{C_{\mathrm{forw}}^{\mathrm{id}}}{K_4 \, (\frac{n_{\mathrm{R}}}{n_{\mathrm{s}}} +1 )\,n_{\mathrm{eq}}^2},\\
    \mathcal{S}_{\mathrm{LU}}^{\mathrm{id}}  &= \frac{C_{\mathrm{forw}}^{\mathrm{id}}}{K_5 \, n_{\mathrm{eq}}^3},
\end{align}
where \(\mathcal{S}_{\mathrm{Jac}}^{\mathrm{id}}\) and \(\mathcal{S}_{\mathrm{LU}}^{\mathrm{id}}\) denote the speedup trends attained if the adjoint runtime were entirely Jacobian- or LU-dominated, respectively. Figure~\ref{fig:esr_speedup} shows that the measured speedup lies close to the Jacobian-dominated bound and remains far below the LU-dominated ceiling. Since the Jacobian dominates the adjoint runtime over the whole range of models considered, the system never enters the LU-dominated regime.
Given the higher computational cost of the Jacobian in the ESR problem, the present implementation may not reach the LU-dominated bound even at the sizes attained in the IDT case, unless extensive, highly implementation-specific optimization is applied, which lies outside the scope of this work.

\bibliographystyle{cnf-num}
\bibliography{cnf-refs}